\documentclass[aps,pre,preprint,superscriptaddress,floatfix]{revtex4-1}
\usepackage{bm}
\usepackage{color}
\usepackage{graphicx}
\usepackage{dcolumn}

\usepackage{times}

\usepackage{amsmath,amssymb}

\begin{document}

\title{Collective animal motion: The case of midge swarms} 
\author{L. L. Bonilla$^*$}
\affiliation{Universidad Carlos III de Madrid, Gregorio Mill\'an Institute for Fluid Dynamics, Nanoscience and Industrial Mathematics, 28911 Legan\'{e}s, Spain}
\affiliation{Universidad Carlos III de Madrid, Department of Mathematics, 28911 Legan\'{e}s, Spain.\\
$^*$Corresponding author. E-mail: bonilla@ing.uc3m.es. ORCID: 000-0002-7687-8595}
\author{R. Gonz\'alez-Albaladejo$^{**}$}
\affiliation{Universidad Carlos III de Madrid, Gregorio Mill\'an Institute for Fluid Dynamics, Nanoscience and Industrial Mathematics, 28911 Legan\'{e}s, Spain.}
\affiliation{Universidad Carlos III de Madrid, Department of Mathematics, 28911 Legan\'{e}s, Spain.\\
$^{**}$ORCID: 0000-0001-9560-5720}
\date{\today}
\begin{abstract}
Among collective states of animal groups, midge swarms pose several puzzling questions. Their polarization order parameter is very small, and the insects interact only weakly with one another while remaining strongly bound to the swarm. In laboratory studies performed without external perturbations, the correlation length is small. By contrast, field studies reveal strong correlations, scale-free behavior, and power-law scaling of the correlation length, susceptibility, and correlation time. The dynamic correlation function exhibits data collapse only for short times when time is rescaled by the correlation time. Is there a theory that explains these disparate observations? We review several models proposed in the literature and extend our harmonically confined Vicsek model to anisotropic confinement. Within the extended criticality region of the phase diagram, numerical simulations predict elongated swarms and static critical exponents that lie between those of the two-dimensional and isotropic three-dimensional models. These values agree better with those measured in natural swarms. We also calculate probability density functions for both natural and laboratory swarms. Numerical simulations indicate that the laboratory distributions correspond to a different region of parameter space, although some open questions remain.
\end{abstract}
\maketitle

\section{Introduction}\label{sec:1}
For animals of such modest size, swarms of midge insects pose a number of puzzling questions. Unlike bird flocks \cite{bal08} and sheep flocks \cite{gin15}, fish schools \cite{her11} or human crowds \cite{kar14}, midge swarms exhibit very weak polarization, and their spatial patterns are comparatively simple \cite{att14plos}. Despite this, natural swarms exhibit strong correlations and scale-free behavior: the correlation length is proportional to swarm size \cite{att14plos,att14}. By contrast, laboratory swarms studied under controlled conditions and in the absence of external perturbations are weakly coupled and exhibit only a short correlation length. It has been argued that external perturbations restore scale-free behavior \cite{vaa20}. Scale-free behavior implies that macroscopic quantities derived from correlation functions obey power laws \cite{att14,cav17}.  

Power laws appear to be ubiquitous in nature \cite{mor11,bia12} but how can be sure they are for real? Stumpf {\em et al.} proposed three criteria for establishing a power law: it has to exhibit a linear behavior in a log-log plot for at least two orders of magnitude, there should be a theory that generates it, and there should be ample and uncontroversial empirical support for it \cite{stu12}. Equilibrium second-order phase transitions (e.g., paramagnetic to ferromagnetic phases at the Curie temperature \cite{cal85,hua87,ami05}) satisfy all these requirements. Wilson's renormalization-group theory has also proved extraordinarily successful in explaining a wide range of critical phenomena beyond its original scope \cite{wil74,wil83,hoh77}. How useful are these ideas to explain midge swarming?

We use the available experimental evidence as a set of benchmarks against which competing theoretical descriptions can be evaluated, rather than reviewing existing models independently. We discuss the extent to which each model accounts for these observations and identify the aspects that remain unexplained. In addition to critical behavior, velocity, position, and acceleration statistics provide independent and stringent tests of competing theories and are therefore included in our assessment. Section \ref{sec:2} summarizes laboratory and field observations. Section \ref{sec:3} reviews four theoretical approaches. Section \ref{sec:4} develops the anisotropic harmonically confined Vicsek model and presents new numerical results for both natural and laboratory swarms. Finally, Section \ref{sec:5} discusses the implications of these results and outlines directions for future work.

 \section{Experiments and observations}\label{sec:2}
\begin{center}
\begin{figure}[ht]
\begin{center}
\includegraphics[clip,width=0.45\linewidth]{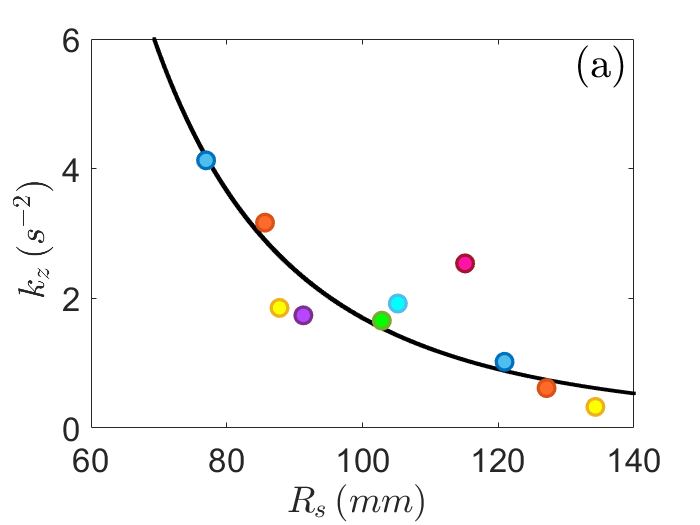}
\includegraphics[clip,width=0.45\linewidth]{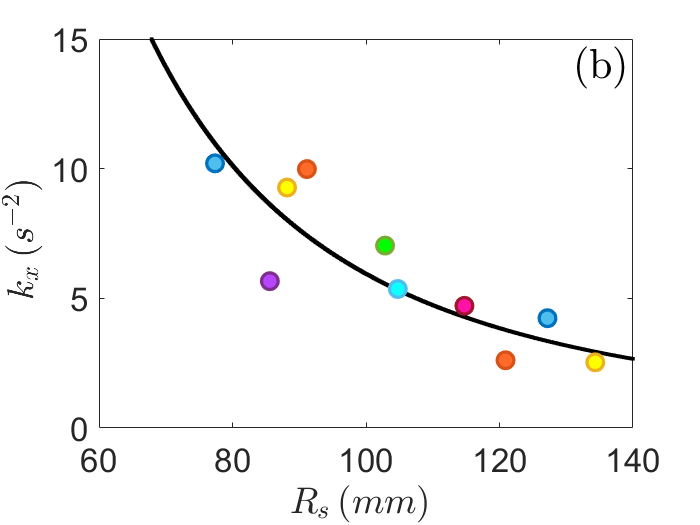}
\end{center}
\caption{Effective spring constants in {\bf (a)} the vertical direction and {\bf (b)} one horizontal direction versus RMS radius for laboratory swarms with $N<100$. Data are extracted from Figure 6 of \cite{kel13} for the mean acceleration conditioned on position with the results: $k_z\sim e^{12.77}R_s^{-(3.4 \pm 0.8)}$, $k_x \sim e^{16.38}R_s^{-(2.4 \pm 0.5)}$. }
 \label{fig1}
\end{figure}
\end{center}

Insect swarming can be observed both in nature and under controlled laboratory conditions. In nature, male midges form swarms at dawn or dusk over distinctive spots on the ground called markers (wet areas, cow dung, man-made objects, etc) \cite{dow55,oku74}. Their purpose is to attract females. Since the 2010s, Ouellette and coworkers have extensively studied swarm formation and properties under controlled laboratory conditions \cite{oue22}. Environmental illumination levels provide a behavioral cue to individual insects for when to swarm. Experiments under controlled illumination show that swarm formation and dissolution are accompanied by the emergence and disappearance of an effective central potential that binds individuals to the swarm \cite{pat22}. Swarms consist of a condensed core and a vapor of insects that leave or enter it \cite{sin17}. Midges acoustically interact when their distances are sufficiently small \cite{att14plos,att14} and react collectively to external acoustic signals \cite{ni15}. The distribution of speeds is peaked about some value and exhibits heavy tails for large swarms (perhaps due to the formation of clusters) \cite{kel13}. The statistics of accelerations of individual midges in a swarm is consistent with postulating a linear spring force (therefore a harmonic potential) that binds insects together \cite{kel13}, which was proposed earlier by Okubo \cite{oku86}. The measured spring constants along the horizontal and vertical directions decrease as the swarm radius increases following power laws; see Fig.~\ref{fig1}, which uses data extracted from \cite{kel13}. 

 For small swarms (50 midges or less), the stationary marginal probability density function (PDF) in space is a $q$-Gaussian ($q>1$) \cite{tsa23}, but it is Gaussian ($q=1$)  for larger swarms (between 20 and 100 midges); see Fig.~1 of \cite{rey21epje}. Within the swarm under laboratory conditions, the correlation length is small, about thrice the midge length \cite{niou15}. These results do not change appreciably if we subtract rotations and dilations from the center of mass at each time \cite{niou15}. The trajectories of single midges follow L\'evy flights \cite{rey16}. Defining effective pressure, temperature and volume of the swarm, its equation of state can be inferred \cite{sin21}, and thermodynamic cycles can be displayed by time varying changes in illumination and acoustic signals \cite{sin21,ni15,sin19}. By moving two markers, one can probe the tensile properties of the swarm and its yield strength \cite{ni16}. A sinusoidal motion of a marker allows measuring the swarm elastic and viscoelastic moduli \cite{vaa19}. The polarization order parameter of the swarm, $W\in[0,1]$: 
\begin{eqnarray}
W(t)= \left|\frac{1}{N}\sum_{j=1}^N \frac{\mathbf{v}_j(t)}{|\mathbf{v}_j(t)|}\right|\!,\quad W=\langle W(t)\rangle_t,\label{eq1}
\end{eqnarray}
 is very small \cite{att14plos,kel13}. Here $\mathbf{v}_j(t)$ is the  velocity of the midge $j$, $j=1,\ldots,N$, at time $t$ and $W$ is the time average over the duration of the film taken from the observation of the swarm.
 
 In nature, midge velocities within swarms are strongly correlated and scale-free, i.e., the correlation length is proportional to the size of the swarm \cite{att14plos,att14,cav17}. In contrast to field studies, midges exhibit only short range velocity correlations in the laboratory (under controlled temperature and no external perturbations): they are on average very weakly coupled, and yet tightly bound to the swarm itself \cite{ni15}. Remarkably, the spaces between insects (defined as centroids of tetrahedra formed by individuals and three nearest neighbors) are strongly correlated and they move as midges in natural swarms \cite{rey23}. Field studies by Cavagna and coworkers have shown that the static and dynamic velocity correlation functions have peculiar qualitative features \cite{cav17}.  In critical dynamics about equilibrium second-order phase transitions \cite{hoh77}, curves of the dynamic correlation function versus time collapse to a single curve as they are written as functions of time scaled with the correlation time (or, equivalently, with correlation length to a power given by the dynamic critical exponent). Field data show that the collapse of the dynamic correlation function with rescaled time occurs only for short times \cite{cav17}. In addition, the dynamic correlation function is flat near the origin when plotted against rescaled time in the sense defined in Ref.~\cite{cav17}. Furthermore, quantities such as the susceptibility, the correlation length and the correlation time follow power laws with characteristic static \cite{att14} and dynamic \cite{cav17,cav23} exponents. 
 
The following definitions will be useful. The static correlation function (SCF) $C(r)$ is  \cite{att14}
\begin{subequations}\label{eq2}
\begin{eqnarray}
&&C(r)= \frac{d Q(r)}{d r},\quad  Q(r)= \left\langle\frac{1}{N}\sum_{i=1}^{N}\sum_{j\neq i}^{N} \delta\hat{\mathbf{v}}_i(t)\!\cdot\!\delta\hat{\mathbf{v}}_j(t)\theta(r-r_{ij}(t))\right\rangle_t, \label{eq2a}\\
&&\delta\mathbf{v}_j(t)=\mathbf{v}_j(t)-\frac{1}{N}\sum_{l=1}^N \mathbf{v}_l(t),\quad \delta\hat{\mathbf{v}}_j(t)=\frac{\delta\mathbf{v}_j(t)}{\sqrt{\frac{1}{N}\sum_{k=1}^N|\delta\mathbf{v}_k(t)|^2}},\quad r_{ij}(t)=|\mathbf{x}_i(t)-\mathbf{x}_j(t)|,\quad\label{eq2b}
\end{eqnarray}\end{subequations}
where $\theta(x)=1$ for $x>0$ and $\theta(x)=0$ otherwise is the unit step function. The insect positions $\mathbf{x}_j(t)$ are measured from the center of mass and $Q(r)$ is the static cumulative correlation function. The Fourier transform of the dynamic correlation function (DCF) is \cite{cav17}
\begin{eqnarray}
\hat{C}(k,t)\!=\!\left\langle\! \frac{1}{N}\sum_{i,j =1}^{N}\!\frac{\sin(kr_{ij}(t_0,t))}{kr_{ij}(t_0,t)}\delta\hat{\mathbf{v}}_i(t_0)\!\cdot\!\delta\hat{\mathbf{v}}_j(t_0+t) \!\right\rangle_{t_0}\!,\quad \langle f\rangle_{t_0}=\frac{1}{t_{max}-t}\sum_{t_0=1}^{t_{max}-t}f(t_0,t),      \label{eq3}
\end{eqnarray}
where $r_{ij}(t_0,t)=|\mathbf{x}_i(t_0)-\mathbf{x}_j(t_0+t)|$, $k$ is the wave number and $t_0$ and $t_{max}$ are the initial and maximum observation time, respectively. From the SCF, the correlation length $\xi=r_0$ and the susceptibility $\chi$ are 
\begin{eqnarray}
C(r_0)=0,\quad \chi=\max_r Q(r)=Q(r_0).\label{eq4}
\end{eqnarray}
 In a second-order equilibrium phase transition, the free energy is a homogeneous function of control parameter (e.g., temperature minus critical temperature), external field and correlation length. Differentiation of the free energy provides power laws for correlation length, susceptibility, etc in terms of the control parameter thanks to the homogeneity property, which is called finite-size scaling (FSS) \cite{ami05}. In field studies of midge swarms, FSS is assumed and the role of the temperature in the control parameter is played by the average distance $x$ between next neighbor insects (rescaled by the body length of a midge), which is called perception range \cite{att14}. Then the following power laws have been established
 \begin{eqnarray}
x-x_c\sim N^{-\frac{1}{3\nu}},\quad\xi\sim N^{1/3}\sim(x-x_c)^{-\nu},\quad\chi\sim(x-x_c)^{-\gamma}\sim\xi^{\gamma/\nu}\sim N^\frac{\gamma}{3\nu}, \label{eq5}
\end{eqnarray}
where the measured static critical exponents are $\nu=0.35\pm 0.1$, $\gamma= 0.9\pm 0.2$. The value of $x$ for finite $N$ is the argument at which the susceptibility reaches its maximum and $x_c=12.5\pm 1.0$ ($N=\infty$). For laboratory swarms, the power laws for the spring constants of Fig.~\ref{fig1} can be rewritten as $R_s\sim k_{z,x}^{-\nu_{z,x}}$ where $\nu_z=0.29$ and $\nu_x=0.42$, which gives an arithmetic mean of $\nu=0.35$. This value is compatible with the previously mentioned static critical exponent of natural swarms if we assimilate $k_{z,x}$ to control parameters (that are zero at the phase transition corresponding to $N=\infty$) instead of $(x-x_c)$ and use the RMS swarm radius $R_s$ instead of $\xi$. 

For critical dynamics about a second-order equilibrium phase transition, the curves $\hat{C}(k,t)$ collapse into a single curve when the time is scaled with the correlation length to the dynamic critical exponent $z$, which is called the {\em dynamic scaling hypothesis} \cite{hoh77}. Assuming this hypothesis to hold, we could write Eq.~\eqref{eq3} as:
\begin{subequations}\label{eq6}
\begin{eqnarray}
&&\frac{\hat{C}(k,t)}{\hat{C}(k,0)}= \left. f\!\left(\frac{t}{\tau_k},k\xi\right)\right|_{k\xi=1}=g(k^zt,k\xi) \quad (k\propto \xi^{-1}),\label{eq6a}\\
&&g(t)=\frac{\hat{C}(k_c,t)}{\hat{C}(k_c,0)};\quad \tau_k=k^{-z}\phi(k\xi). \label{eq6b}
\end{eqnarray}
Here $\tau_k=k^{-z}\phi(k\xi)$ is the correlation time at wave number $k$, found by solving the equation \cite{cav17,gon23,gon24}:
\begin{eqnarray}
\sum_{t=0}^{t_{max}} \frac{1}{t}\,\sin\!\left(\frac{t}{\tau_k}\right) f\!\left(\frac{t}{\tau_k},k\xi\right)\! = \frac{\pi}{4}.  \label{eq6c}
\end{eqnarray}
\end{subequations}
The value at which $\hat{C}(k,t)$ reaches its maximum, $k_c$, is inversely proportional to the correlation length (so that we can redefine $\xi=1/k_c$, hence $\tau_{k_c}= \phi(1) \xi^z$). Note that the perception range is not necessary to calculate the dynamic critical exponent $z$. However, its value depends on the  method used to fit the data. Least square (LS) fitting yields $z=1.16\pm 0.12$ whereas reduced major axis (RMA) fitting produces $z=1.37\pm 0.11$ \cite{cav23}. With data obtained from field studies, the collapse into a single function $g(x)$ as given by Eq.~\eqref{eq6a} does not occur: All $\hat{C}(k,t)$ curves collapse into a single one but only on the interval $0<k_c^zt<4$ and they clearly do not overlap for larger values of $k_c^zt$ (cf Fig.~2 of Ref.~\onlinecite{cav17}). Besides partial collapse of the DCF, the function $-x^{-1}\ln g(x)$ tends to 0 (it is flat) as $x=k_c^zt\to 0$ \cite{cav17}, which is the other qualitative feature mentioned before. These two qualitative features are also present in data from laboratory swarms when the DCF is calculated from the moving spaces between midges and not from the weakly correlated insect themselves; see Fig. 5 of \cite{rey23} (data from $N=15,46,94$).
 
 It is interesting to note that the data used to get the power laws and critical exponents for natural swarms were collected in different days over many years from swarms having different number of insects, and there were swarms of different midge species ($N$ ranges from 69 to 781; see Supplementary Materials of Refs.~\onlinecite{att14,cav17,cav23}).
 
 The experimental observations summarized above provide the benchmarks against which the theoretical models discussed in the next section will be assessed.
 
 \section{Theories}\label{sec:3}
Here we present different theories of swarm formation and discuss whether they explain the observations described in Section \ref{sec:2}. The following experimental observations provide the principal criteria for assessing the different theories. For natural swarms, the existence of power laws, small polarization but strong velocity correlations are important. For laboratory swarms, small correlation length and effective spring constants are important. The available theoretical descriptions emphasize different aspects of the experimental phenomenology. Models developed to explain natural swarms have primarily focused on critical behavior and scaling properties, whereas those proposed for laboratory swarms have concentrated on stationary statistical properties and effective single-particle descriptions. For this reason, different models are naturally assessed using different subsets of the available experimental evidence. Which framework ultimately explains the largest body of independent observations?

 \subsection{Periodic Vicsek model}
The periodic Vicsek model (PVM) \cite{vic95,vic12} exhibits an ordering transition that could, in principle, provide a description of the critical behavior observed in natural swarms. It is defined by the equations
\begin{eqnarray}
&&\mathbf{x}_j(t+\Delta t)=\mathbf{x}_j(t)+ \Delta t\,\mathbf{v}_j(t+\Delta t),\quad j=1,\ldots,N,\nonumber\\
&& \mathbf{v}_j(t+\Delta t)=v  \mathcal{R}_\eta\!\left[\Theta\!\left(\sum_{|\mathbf{x}_l-\mathbf{x}_j|<r_1 R_0}\mathbf{v}_l(t)\right)\!\right]\!. \label{eq7}
\end{eqnarray}
Here $\Theta(\mathbf{x})=\mathbf{x}/|\mathbf{x}|$, $R_0$ is the radius of the sphere of influence about particles, $v_0$ is the constant particle speed, and $\mathcal{R}_\eta(\mathbf{w})$ performs a random rotation uniformly distributed on a spherical sector around $\mathbf{w}$ with maximum opening $\eta$ \cite{att14}. Particles align their velocities with the mean of their neighbors within a sphere of influence except for an alignment noise of strength $\eta$ and are confined to a cube of finite volume by periodic boundary conditions. The noise and the extension of the influence sphere are presumably due to uncontrollable external disturbances in the wild. As $v\Delta t/r_1$ tend to zero, the discrete time equations \eqref{eq7} become differential equations that describe point particles undergoing overdamped dynamics with forces that tend to align their velocities. The  time step in the VM can be interpreted as the time insects take to be aware of their neighbors and decide to align with the mean direction they provide. In nondimensional units, the discrete time step is not small although the real motion of the insects occurs in continuous time.

In 2D, the PVM exhibits an ordering phase transition from a spatially homogeneous phase (zero polarization, $W=0$) to traveling bands ($W>0$) \cite{cha20}. For a sufficiently small box length (much smaller than the period of the traveling bands), this ordering transition is continuous. Thus it would seem that this model is a plausible candidate to explain swarm formation if studied on the disordered phase near the phase transition (leaving aside the fact that observed swarms are far from being spatially homogeneous). However, Cavagna {\em et al} have shown that this is not the case: The DCF $g(k_c^zt)$ is not flat at the origin ($-x^{-1}\ln g(x)\to 1$ as $x\to 0$), and the model produces power laws whose numerically calculated critical exponents are far from the measured ones (e.g., $z=2$, far from 1.37) \cite{cav17}. Overall, the PVM provides a valuable starting point for understanding collective alignment and ordering transitions. However, in its original form it does not account for the observed critical exponents or the qualitative features of the dynamic correlation function (DCF).

\subsection{Inertial spin models}
The inertial spin model (ISM) was introduced to account for the nearly flat behavior of the dynamic correlation function (DCF) at very small scaled times. It is defined by the equations \cite{cav15,cav24}
\begin{eqnarray}
&&\frac{d\mathbf{x}_j}{dt}=\mathbf{v}_j,\quad j=1,\ldots,N,\nonumber\\
&&\frac{d\mathbf{v}_j}{dt}=\frac{1}{\chi}\mathbf{s}_j\times \mathbf{v}_j,\nonumber\\
&&\frac{d\mathbf{s}_j}{dt} =\mathbf{v}_j\times \frac{J}{n_i} \sum_{l=1}^N n_{jl}(t) \mathbf{v}_l -\frac{\eta}{\chi}\mathbf{s}_j - \mathbf{v}_j\times \boldsymbol{\zeta}_j, \label{eq8}
\end{eqnarray}
to be solved with periodic boundary conditions. Here the zero-mean white noise has correlation $\langle \zeta^\mu_j(t) \zeta^\nu_l(t')\rangle = 2T \eta\delta_{jl}\delta_{\mu\nu}\delta(t - t')$, $n_{jl}=\theta(R_0-|\mathbf{x}_j-\mathbf{x}_l|)$, $n_j=\sum_{l=1}^Nn_{jl}$. The parameter $J$ is the alignment strength, $\eta$ is the microscopic spin dissipation, $\chi$ is a spin scale, and $T$ is the noise amplitude (or temperature). Clearly the speed of the particles is conserved by these equations so that we can impose that they all have the same speed $v_0$. In the overdamped limit, we may ignore $d\mathbf{s}_j/dt$ and obtain the following equation:
\begin{eqnarray*}
\eta\frac{d\mathbf{v}_j}{dt}=\!\left(\mathbf{v}_j\times \tilde{J} \sum_{l=1}^N n_{jl}(t) \mathbf{v}_l +\boldsymbol{\zeta}_j\right)\!\times\mathbf{v}_j,
\end{eqnarray*}
with $\tilde{J}$ instead of $J/n_j$. This is a version of the continuous time Vicsek model and it has an ordering transition. Eq.~\eqref{eq8} adds inertia so that the resulting ISM may produce a DCF which is flat at the origin near the ordering transition \cite{cav15,cav24}. 

It turns out that the periodic ISM near its ordering transition produces the same dynamic critical exponent as the active irreversible version of model G in Ref.~\cite{hoh77}. The resulting model nonlinearly couples a field of unit velocity vectors with a spin field and it includes space-time white noise sources. At the ordering transition, Cavagna {\em et al} have found the renormalization group equations (RGEs) of the active G model within the one-loop approximation by using the $4-d$ Wilson expansion \cite{cav23}. The authors consider that one of the four critical points of the RGEs corresponds to the behavior of midge swarms and proceed to find the critical exponents using the most appropriate fixed point with the result: $z=1.35$, $\nu = 0.748$, $\gamma = 1.171$. Numerical solution of the ISM produces the same exponent $z$ and so it does a version of the ISM (fully-conserved ISM) that is closer to the active model G \cite{cav24}. Thus, at their ordering phase transition, the active G model and the different ISMs may belong to the same universality class in Wilson's sense \cite{wil74}.

While the dynamical critical exponent predicted by these theories is close to observations using the RMA fit \cite{cav23}, the predicted static exponents are far from those measured ($\nu=0.35$, $\gamma=0.9$) \cite{att14}. These models also produce a DCF which is flat near zero time. However, the data collapse of the DCF with scaled time $t/\xi^z$ occurs for all times, not for the finite time interval observed in natural swarms \cite{cav17}. In addition, the qualitative shape of the observed swarms is quite different from the uniform density phase expected near an ordering transition of the ISMs. Thus, the ISM theories at their ordering transition do not describe midge swarming observed in field studies. Overall, the inertial spin model successfully reproduces the observed dynamical critical exponent and its ordering transition belongs to the same universality class as active Model G. However, it does not account for the experimentally observed static critical exponents or the partial data collapse of the DCF, which is observed only at small scaled times.

\subsection{Effective one-particle theories}
Swarms in laboratory experiments consist of midges that are on average very weakly coupled, and yet tightly bound to the swarm itself \cite{ni15}. Consequently, Ouellette and coworkers have proposed effective one-particle ({\em mean-field}) theories: one representative particle moves within the swarm force field produced by all the particles. While the mean field in statistical mechanics is formally obtained from the Bogoliubov inequality for the canonical distribution \cite{cal85}, other considerations have to be made for swarms. Effective one-particle theories have undergone substantial development over the past decade. Since a comprehensive review is beyond the scope of this article, we focus on the formulations most directly related to the experimental observations discussed in Section \ref{sec:2}. Consider for example the probability density of one particle in a swarm velocity-dependent potential satisfying the Fokker-Planck equation (FPE) \cite{rey17,vaa19}
\begin{subequations}\label{eq9}
\begin{eqnarray}
&&\frac{\partial P}{\partial t}+\mathbf{v}\cdot\nabla_xP +\nabla_v\cdot\!\left[\left(-\frac{\mathbf{v}}{T} + \langle \mathbf{A}|\mathbf{v},\mathbf{x}\rangle\right)\! P\right]\! = \frac{\sigma^2_v}{T}\nabla_v^2P, \label{eq9a}\\
&& \langle\mathbf{A}|\mathbf{v},\mathbf{x}\rangle= -\frac{\mathbf{v}\cdot(\mathbf{x}-\mathbf{x}_c)}{\sigma_r^2 |\mathbf{v}|^2} \mathbf{v}\times \left\{\begin{array}{cc} 3\sigma_v^2 &\mbox{(Gaussian velocities)}\\ 2\langle s\rangle(s+\langle s\rangle) &\mbox{(exponentially distributed)}
\end{array}\right.,\quad s=|\mathbf{v}|, \label{eq9b}\\
&&P_s(\mathbf{x},\mathbf{v})=\frac{1}{(2\pi\sigma_r^2)^{3/2}}\exp\!\left[-\frac{|\mathbf{x}-\mathbf{x}_c|^2}{2\sigma_r^2}\right]\!\times\! \left\{\begin{array}{cc} \frac{e^{-\frac{s^2}{2\sigma_v^2}}}{(2\pi\sigma_v^2)^{3/2}} &\mbox{(Gaussian velocities)}\\ \frac{e^{-s/\langle s\rangle}}{\langle s\rangle} &\mbox{(exponentially distributed)}
\end{array}\right.. \label{eq9c}
\end{eqnarray}
Here $\langle\mathbf{A}|\mathbf{v},\mathbf{x}\rangle$ and $\langle s\rangle$ are the conditional mean acceleration and the mean speed, respectively. The conditional mean acceleration expresses the effective restoring force that binds individuals to the swarm whose center is $\mathbf{x}_c$. It is determined by assuming that the stationary probability density of Eq.~\eqref{eq9c} is Gaussian in space and either Gaussian or exponentially distributed in speed \cite{rey17}. The speed statistics are radially independent (homogeneous), as observed in lab swarms \cite{rey17}. The stationary distributions are maximum-entropy distributions \cite{jay57,set06} subject to experimentally motivated constraints \cite{rey17,kel13}, and are therefore minimally biased statistical descriptions. See also \cite{rey24,rey24entropy}. The free parameters $T$, $\sigma_r$ (root-mean-square, RMS, swarm radius) and $\sigma_v$ (RMS swarm velocity) are set to unity. For Gaussian velocities, the mean acceleration (i.e., the restoring force) increases linearly with the distance to the swarm center, which is reminiscent of Okubo's model \cite{oku86} and of Gorbonos {\em et al}'s self-gravitating model \cite{gor16}. 
\end{subequations}

The Langevin equations corresponding to Eqs.~\eqref{eq9} are
\begin{subequations}\label{eq10}
\begin{eqnarray}
&&d\mathbf{x}=\mathbf{v}\, dt,  \label{eq10a}\\
&& d\mathbf{v}=\!\left(\langle\mathbf{A}|\mathbf{v},\mathbf{x}\rangle-\frac{\mathbf{v}}{T}\right) dt+ \sqrt{\frac{2\sigma_v^2}{T}}\, d\mathbf{W}(t), \label{eq10b}
\end{eqnarray}\end{subequations}
where $\mathbf{W}(t)$ is the usual Wiener process. Solving Eqs.~\eqref{eq10} for many independent realizations of the stochastic process mimics simulating many particles within a swarm that have zero correlation among them. For exponentially distributed $P_s$, $\langle s\rangle$ can be either the constant mean of Eq.~\eqref{eq9c} or the time-dependent mean speed, in whose case the FPE Eq.~\eqref{eq9a} is nonlinear in $P$. For undisturbed swarms $\mathbf{x}_c$ can be the origin of coordinates. 

Experiments that probe the equation of state and thermodynamic cycles \cite{sin21}, tensile properties \cite{ni16}, or viscoelasticity \cite{vaa19} require moving markers, thereby posing a time dependent $\mathbf{x}_c$. Assuming a motion $\mathbf{x}_c(t)$ (for example, a periodic oscillation of the marker about the origin with $\mathbf{x}(0)=\mathbf{0}$), disturbances in the center of mass can be modeled  as a damped shear wave propagating in the vertical direction \cite{vaa19}. Then the elastic storage and viscoelastic loss moduli can be calculated and compared with data \cite{vaa19}. van der Vaart {\em et al} demonstrate convincingly that external perturbations can generate long-range correlations in laboratory swarms \cite{vaa20}. Their observation does not establish that the perturbed swarm has entered a critical regime. In their stochastic model based on Eqs.~\eqref{eq10}, the scaling argument identifies the RMS swarm size with the correlation length, whereas the latter should be obtained from a two-point correlation function. The authors of \cite{vaa20} perform simulations of $N$ replicas of Eqs.~\eqref{eq10} with a $\mathbf{x}_c(0)$ displaced from the origin. The results are a SCF and a cumulative correlation function resembling those of natural swarms, and a susceptibility that increases linearly with $N$ \cite{vaa20}. Furthermore, by replacing $\sigma_r$ by an adjustable RMS swarm size that serves as the correlation length (but it has not been calculated from a two point correlation function), it is possible to argue that disturbances in the CM  yield data collapse of the DCF with dynamic critical exponent $z=1$ \cite{vaa20}. While the swarm radius is used as a proxy for the correlation length, these quantities are conceptually distinct. A system can have $R_s\sim N^{1/3}$ simply because more independent particles are put into a larger cloud. That does not imply $\xi\sim N^{1/3}$. The latter is a statement about the decay of a two-point correlation function. Correlations induced by a common external perturbation are not equivalent to intrinsic many-body correlations generated by interparticle interactions, and their existence does not establish criticality.  

While midges under lab conditions are weakly correlated, the spaces between them are strongly correlated as in natural swarms \cite{rey23}. Simulations of 100 replicas (non-interacting individuals) of Eqs.~\eqref{eq10} with $\sigma_v=T=1$ and $\sigma_r$ between 1 and 1/4 (varying RMS swarm size) show the following. Spaces between individuals exhibit data collapse of the DCF in terms of $k^zt$ with $z=1$ (Figures 3 and 6). Contrastingly with the simulations, spaces within midges in lab data show only partial collapse of the DCF with scaled time; see Fig. 5 of \cite{rey23}.

For quiescent laboratory conditions, effective one-particle models {\em different} from Eqs.~\eqref{eq9} and \eqref{eq10} predict correctly the emergence of macroscopic mechanical properties. Replacing Eq.~\eqref{eq9b} by the 1D restoring force of a bimodal swarm, $\sigma_v^2 (ax-bx^3)$ (where $x=\pm. \sqrt{a/b}$ locate markers being pulled apart), and adding multiplicative noise of strength $x$, Reynolds has argued that swarms have a Young modulus and a yield  strength \cite{rey19}, as observed in \cite{ni16}. To predict the equation of state and thermodynamic cycling, Reynolds uses a 1D version of Eqs.~\eqref{eq10}, in which the acceleration is replaced by the derivatives to a linear combination of Gaussian functions (in the heights $z-\overline{z}_j$, $j=1,2$) representing slabs within the swarm plus multiplicative noises with the same strengths \cite{rey21}. The thermodynamic cycle is obtained by cycling the additive and multiplicative noises \cite{rey21}. Note that the forces are now the derivatives of Gaussian potentials (not linear springs). 

In summary, effective one-particle theories describe midges in a swarm as independent particles moving in a mean-field potential derived from a stationary probability density, e.g., Eq.~\eqref{eq9c}. Different stationary probability densities correspond to minimally biased statistical descriptions subject to experimentally motivated constraints. The different variants reviewed above illustrate the versatility of the effective one-particle approach and the range of laboratory phenomena it can successfully describe. The calculation of the dynamic critical exponent relies on identifying the swarm size with the correlation length. However, particles corresponding to different realizations of the Langevin equations \eqref{eq10} are statistically independent. In fact, static critical exponents have not been calculated within the effective one-particle theories \cite{vaa20} and therefore they cannot be compared to static critical exponents measured in natural swarms \cite{att14}. Furthermore, the underlying stationary probability density may depend on the particular experiment being modeled. Would it be possible to find a unifying  theory for all observations?

To go beyond mean field, understand its applicability, and obtain critical exponents comparable with the observations, a ``microscopic theory'' specifying the interaction between midges is required. Such a theory should also explain the difference between observations of small laboratory swarms and larger natural swarms. At present, effective one-particle theories do not yield the experimentally observed critical exponents of natural swarms.

\subsection{Isotropic harmonically confined Vicsek model}
The VM may move aligned particles away to the borders of a finite box, which is avoided by imposing periodic boundary conditions. However, swarms observed in the laboratory seem to adopt their shape by experiencing a central force that pulls them together, as explained in Section \ref{sec:2}. Adding a spring force to Eq.~\eqref{eq7}, we obtain the harmonically confined Vicsek model (HCVM):
\begin{eqnarray}
&&\mathbf{x}_j(t+\Delta t)=\mathbf{x}_j(t)+ \Delta t\,\mathbf{v}_j(t+\Delta t),\quad j=1,\ldots,N,\nonumber\\
&& \mathbf{v}_j(t+\Delta t)=v  \mathcal{R}_\eta\!\left[\Theta\!\left(\sum_{|\mathbf{x}_l-\mathbf{x}_j|<r_1 R_0}\mathbf{v}_l(t)-\mathcal{B} \mathbf{x}_j\right)\!\right]\!. \label{eq11}
\end{eqnarray}
Here $\mathcal{B}$ is a diagonal matrix of positive spring constants, and positions and velocities take values on $\mathbb{R}^3$. In our previous works, we considered an isotropic matrix with spring constant $\beta_0$ \cite{gon23,gon24}. In effective one-particle theories, the probability density is postulated first, then the force follows. In contrast, the HCVM starts from the microscopic equations of motion \eqref{eq11}, and the macroscopic properties emerge from their dynamics. Here we consider the anisotropic case, in which the spring constant along the vertical axis is smaller than those along the horizontal axes.

The nondimensional version of Eq.~(\ref{eq11}) has $\Delta t=1$, $r_1=1$, speed $v_0$ and confinement parameter $\beta$ given by
\begin{eqnarray}
v_0=v\, \frac{\Delta t}{r_1}, \quad \beta=\beta_0 \Delta t. \label{eq12}
\end{eqnarray} 
From  observations of natural midges reported in the supplementary material of Refs.~\onlinecite{cav17,att14}, $v_0$ is of order unity \cite{gon23}, which is far from the continuum limit $v_0\ll 1$ at which time differences become time derivatives. Eqs.~\eqref{eq11} have different attractors, periodic, quasiperiodic and chaotic, influenced by noise. The chaotic attractors are characterized by a positive largest Lyapunov exponent (LLE) that typically increases with noise strength \cite{gon23,gon24}. Fig.~\ref{fig2}(a) from Ref.~\cite{gon24} (reproduced below) shows the phase diagram of the HCVM on the confinement vs noise $(\eta,\beta)$ plane for $N=500$, $v_0=R_0=1$. There are regions of deterministic and noisy chaos, noisy period-$\sigma$ (NP$\sigma$, not shown), noisy quasiperiodic (NPQ) attractors, and mostly noise. See Ref.~\cite{gon23} for the technical definition of noisy chaos using scale-dependent Lyapunov exponents, calculations of the LLE and reconstruction of chaotic attractors from time series obtained from the numerical simulations of the HCVM.  

\begin{widetext}
\begin{center}
\begin{figure}[ht]
\begin{center}
\includegraphics[clip,width=16cm]{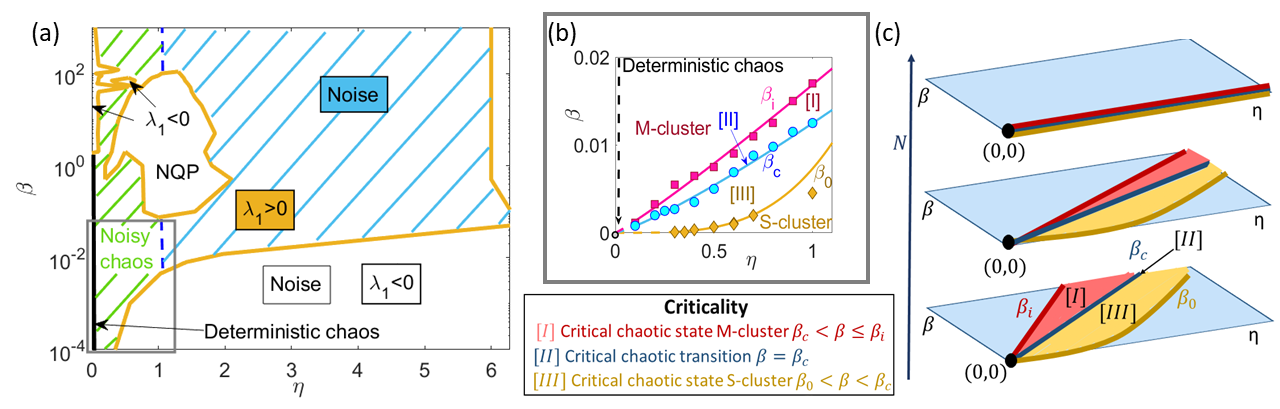}\\
\end{center}
\caption{Isotropic confinement. {\bf (a)} Phase diagram of the confinement vs noise plane indicating regions of deterministic and noisy chaos, noisy quasiperiodic (NPQ) attractors, and mostly noise. {\bf (b)} Regions [I]  $(\beta_c(\eta;N),\beta_i(\eta;N))$ (M-cluster or multicluster chaos), [III] $(\beta_0(\eta;N),\beta_c(\eta;N))$ (S-cluster or single-cluster chaos), and line [II] $\beta=\beta_c(\eta;N)$. In (a) and (b), $N=500$. {\bf (c)} Shrinking of the criticality region as $N$ increases. {\em Adapted from Fig.~1 of Ref.~\onlinecite{gon24}}. 
 \label{fig2}}
\end{figure}
\end{center}
\end{widetext}

The curve $\beta_0(N;\eta)$ separates chaotic and nonchaotic attractors for fixed $N$. On this curve, the LLE is zero and the correlation length is proportional to the size of the swarm. This indicates scale-free behavior and characterizes the scale-free-chaos phase transition in the limit as $N\to\infty$. Fig.~\ref{fig2}(b) shows different critical curves in the phase diagram. Besides $\beta_0(N;\eta)$, the correlation time reaches its minimum for fixed $N$ and $\eta$ at the curve $\beta_c(N;\eta)$. This curve separates the region of chaotic attractors where all particles belong to a single cluster from the region of multicluster chaotic attractors \cite{gon23}. The LLE decays to zero as $\lambda_1\sim\beta_c(N;\eta)^{\varphi}$, $\varphi=z\nu$, as $N\to\infty$ \cite{gon23}. The LLE is maximal on the curve $\beta_i(N;\eta)$. As $N\to\infty$ these three critical curves collapse to $\beta=0$ at the same rate as depicted in Fig.~\ref{fig2}(c) and the time averaged polarization also goes to zero. The line $\beta=0$ for a range of positive $\eta$ is the scale-free chaos phase transition at $N=\infty$ responsible for the critical static and dynamic exponents of natural swarms. Intriguingly, lab swarms have a small correlation length \cite{kel13}, but the swarm radius (a proxy of $\xi$ as $N\to\infty$) goes to infinity as confinement tends to zero; cf Fig.~\ref{fig1}.

For finite $N$, there is a {\em region of extended criticality} between the curves $\beta_0(N;\eta)$ and $\beta_i(N;\eta)$ \cite{gon24}. The correlation length and susceptibility are $\xi=r_0=$ argmax$Q(r)$ and $\chi=$ max$Q(r)$, respectively, where $Q(r)$ is the static cumulative correlation function, cf Eq.~\eqref{eq4}. These quantities can be related to $\beta_0$ and $\beta_c$ for different values of $N$ and the noise $\eta$, which produce the power laws $r_{0j}\propto\beta_j^{-\nu}$ and $\chi\propto\beta_j^{-\gamma}$. The same static critical exponents $\nu$ and $\gamma$ can be obtaining by studying the limit as $\eta\to 0$, in which the following relations hold \cite{gon24}:
\begin{subequations}\label{eq13}
\begin{eqnarray}
&& r_{0j}=D_j \beta_j^{-\nu} \eta^{-p_j}=\frac{D_j }{C_j^\nu}N^\frac{1}{3} \eta^{-p_j - \nu m_j}, \quad j=0,c,\quad \label{eq13a}\\
&&\chi = Q(r_0)=  Q_j\beta_j^{-\gamma}\eta^{q_j} =  \frac{Q_j}{C_j^\gamma}N^\frac{\gamma}{3\nu}\eta^{q_j-\gamma m_j},     \label{eq13b}
\end{eqnarray}
Eqs.~\eqref{eq13a} and \eqref{eq13b} allows to calculate the static critical exponents by fixing $N$ and drawing the curves $\beta_j(N;\eta)$ ($j=0,c$) as functions of $\eta$ (for small values of $\eta$). Furthermore, there is a linear relation between rescaled versions of  the perception range $x$ and $\beta_0$ or $\beta_c$:
\begin{eqnarray}
&&\eta^{m_{x0}}x=A_0+B_0\beta_0\eta^{-m_0},\quad\mbox{i.e., } \beta_0=B_0^{-1} \eta^{m_0+m_{0x}} (x-x_c),\label{eq13c}\\
&&\eta^{m_{xc}}x=A_c-B_c\beta_c\eta^{-m_c}, \quad\mbox{i.e., } \beta_c=B_c^{-1} \eta^{m_c+m_{cx}}(x_c-x).\label{eq13d}
\end{eqnarray}
\end{subequations}
Here the critical perception range at zero confinement is $x_c(\eta)=A_j\eta^{-m_{xj}}$, $j=0,c$. See Ref.~\onlinecite{gon24} for parameter values. Eqs.~\eqref{eq13c} and \eqref{eq13d} show that the critical confinements are proportional to $|x-x_c|$ for a fixed noise strength. Then the static power laws can be written in terms of  $|x-x_c|$ as in Eq.~\eqref{eq5}. The link between the (confinement) control parameter of the phase transition and the perception range allows us to relate theory and measurements of natural swarms. It was found for the first time in Ref.~\onlinecite{gon24}.

For the 3D isotropic HCVM, the calculated critical exponents are $\nu=0.43\pm 0.03$, $\gamma = 0.92\pm 0.13$, and $z=1.24\pm 0.11$ (LS fit) or $z=1.37\pm 0.10$ (RMA fit) \cite{gon24}. The susceptibility as $\eta\to 0$ is calculated in linear response theory; see Eqs. (9)-(11) in Ref.~\onlinecite{gon24}. To obtain the dynamic critical exponents, we proceed as follows. Recall that data from natural swarms correspond to different times, days or years, insect numbers and even midge species \cite{att14,cav17,cav23}. Thus, data from natural swarms are a mixture belonging to the extended criticality region where power laws hold. For numerical simulations, we choose different values of $N$ and $\eta$ on the curves $\beta_0$ and $\beta_c$. This data mixture produced different values of $z$ when depending on whether the fitting is linear squares or reduced major axis.

It is interesting to compare these critical exponents with those given by the mean field version of the HCVM. The mean field equations are \eqref{eq11} for $N=1$ and isotropic confinement \cite{gon23mf}. The resulting critical exponents are those of the Landau theory of equilibrium phase transitions \cite{hua87}, $\nu= 0.5$, $\gamma=1$, with dynamic exponent $z=1$ \cite{gon23mf}.

Numerical simulations have shown that the isotropic HCVM undergoes a scale-free chaos phase transition at zero confinement over a finite noise range in the thermodynamic limit ($N=\infty$) \cite{gon23,gon24}. Although the predicted value of $\nu$ is slightly larger than the experimental estimate, it remains consistent with the reported uncertainty. The isotropic HCVM also reproduces several qualitative features of the observations. It predicts a partial collapse of the DCF when plotted against the scaled time $t/\xi^z$ and a flat behavior as $t/\xi^z\to 0$. The simulated swarms are approximately spherical, whereas natural swarms are vertically elongated. Other features shared with laboratory swarms include a condensed core surrounded by a vapor of insects that continuously enter and leave it \cite{sin17}, together with the disordered motion of both individual insects and the swarm center of mass \cite{rey16}. Overall, the HCVM currently accounts for the broadest range of experimental observations in natural swarms among the theoretical approaches reviewed here. However, it has not yet been developed to explain the full range of observations obtained in laboratory swarms. In particular, the difference between the correlation lengths observed in laboratory and natural swarms remains unexplained.

\section{Anisotropic harmonically confined Vicsek model}\label{sec:4}
Previous work showed that isotropic confinement produces approximately spherical swarms, whereas natural swarms are vertically elongated. Motivated by this discrepancy, we introduce anisotropic confinement in Eq.~\eqref{eq11}:
\begin{eqnarray}
\mathcal{B}_{11}=\mathcal{B}_{22}=\beta, \quad\mathcal{B}_{33}=\frac{\beta}{2}.\label{eq14}
\end{eqnarray}
These values of the spring constants are inspired by Figure 1a of Ref.~\cite{att14} that shows a swarm twice as tall as it is wide.

\begin{center}
\begin{figure}[ht]
\begin{center}
\includegraphics[clip,width=0.9\linewidth]{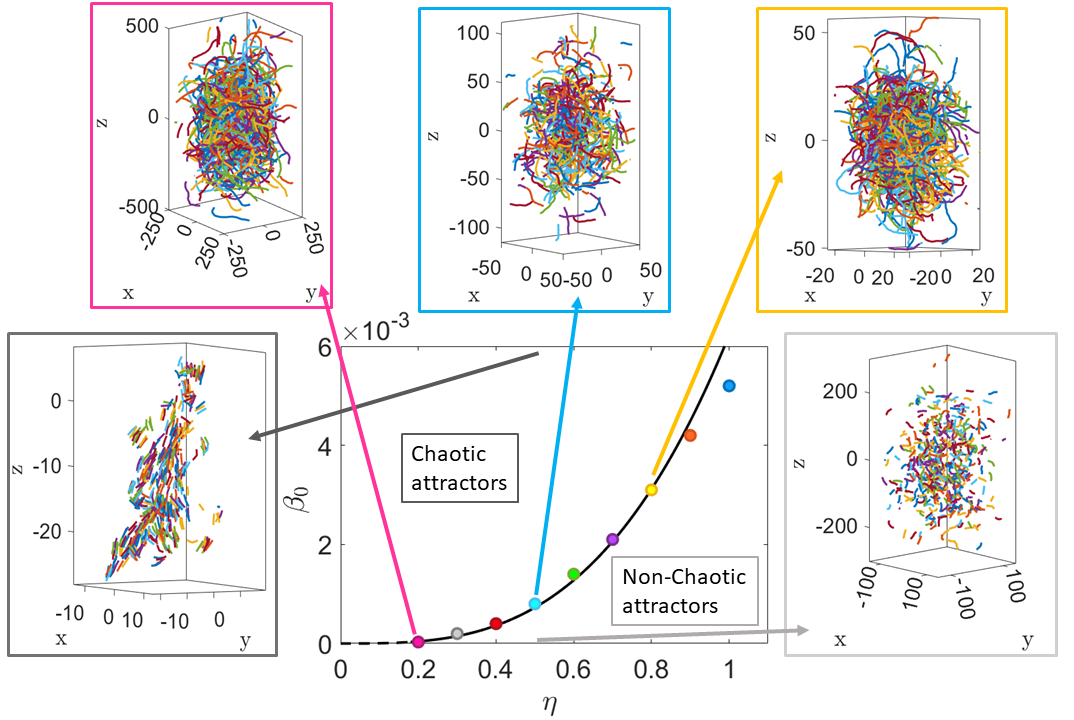}\\
\end{center}
\caption{Phase diagram of the anisotropic HCVM in the $(\eta,\beta)$ plane for $N=500$. The chaotic and nonchaotic regions are separated by the critical curve $\beta_0(\eta)$, along which the largest Lyapunov exponent vanishes. The solid curve is the fit  $\beta_0 = c_0 \eta^{m_0}$, with $m_0 = 3.10 \pm 0.13$ and $c_0 = 0.006 \pm 0.001$. Representative swarm configurations illustrate the evolution of the swarm morphology across the phase diagram.}
 \label{fig3}
\end{figure}
\end{center}

Previous work showed that isotropic confinement produces approximately spherical swarms, whereas natural swarms are vertically elongated. We now investigate whether anisotropic confinement yields more realistic swarm shapes while preserving the critical behavior. To this end, we first compute the critical curve $\beta_0(500;\eta)$ separating the chaotic and nonchaotic regions of the phase diagram. The resulting phase diagram is shown in Figure \ref{fig3}, where the largest Lyapunov exponent vanishes along the critical curve. Figure \ref{fig3} also shows representative short-time trajectories of the midges for different confinement values on the critical curve, together with one swarm deep inside the chaotic region and another in the nonchaotic region. Along the critical curve the swarms are vertically elongated, whereas particles outside the core lie farther from it in the nonchaotic region. Deep inside the chaotic region, the onset of multicluster chaos becomes apparent. The swarm occupies a progressively smaller volume as the confinement increases. Video 1 in the Supplementary Material illustrates the particle dynamics for $\eta=0.5$ at critical confinement.

\begin{center}
\begin{figure}[ht]
\begin{center}
\includegraphics[clip,width=0.3\linewidth]{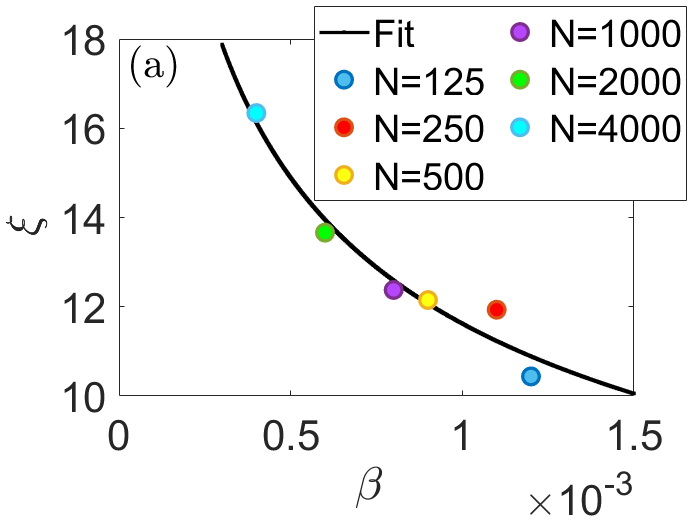}
\includegraphics[clip,width=0.3\linewidth]{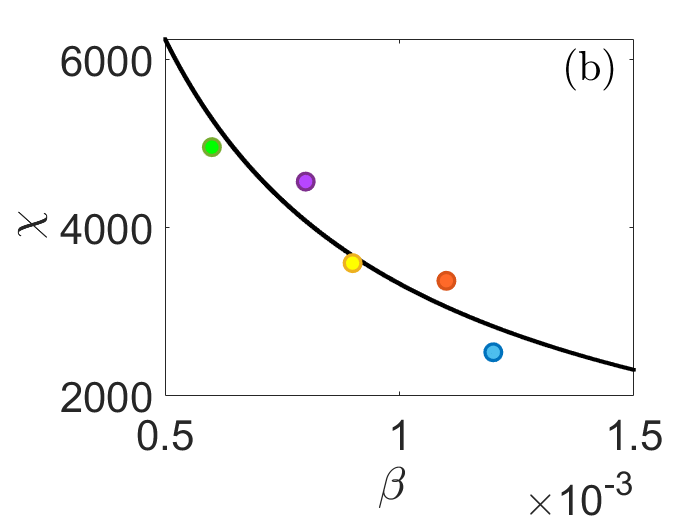}
\includegraphics[clip,width=0.3\linewidth]{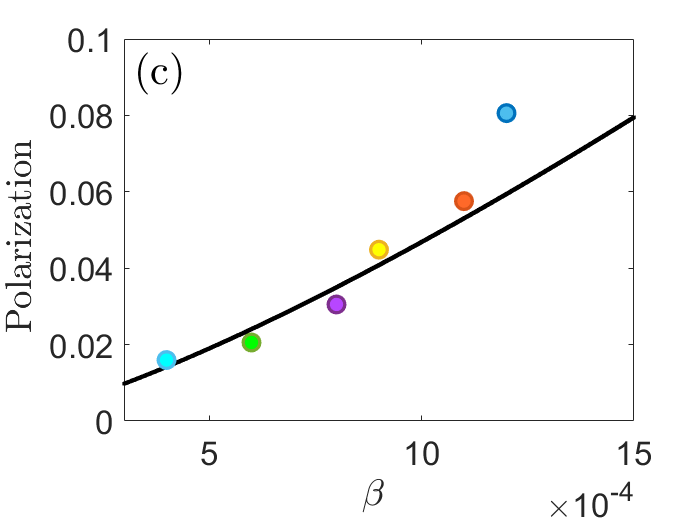}
\includegraphics[clip,width=0.3\linewidth]{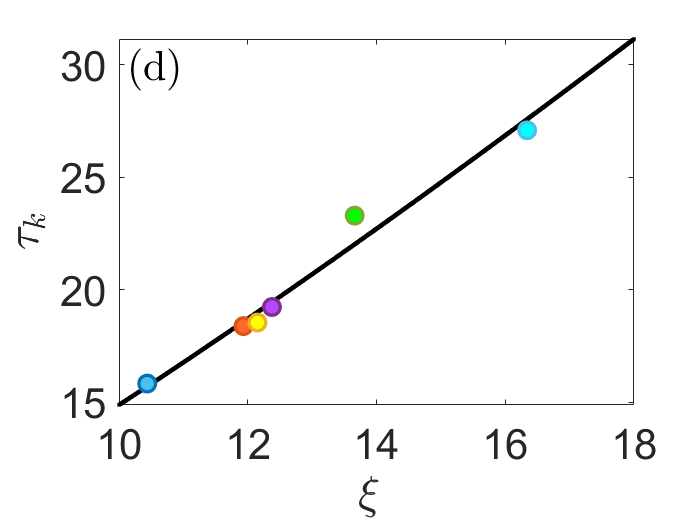}
\includegraphics[clip,width=0.3\linewidth]{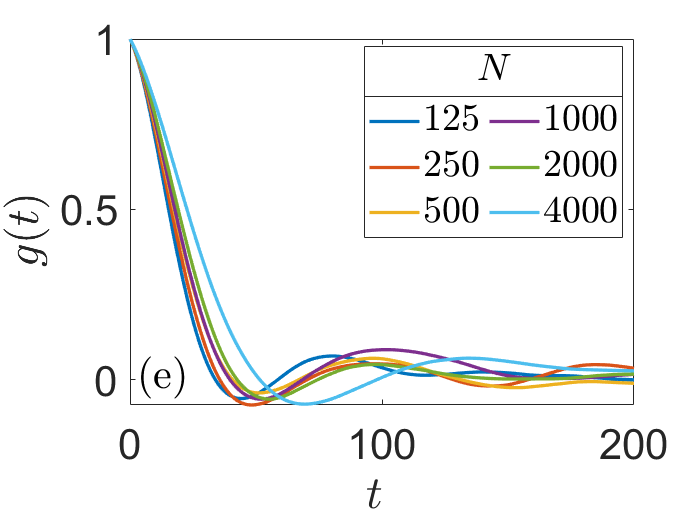}
\includegraphics[clip,width=0.3\linewidth]{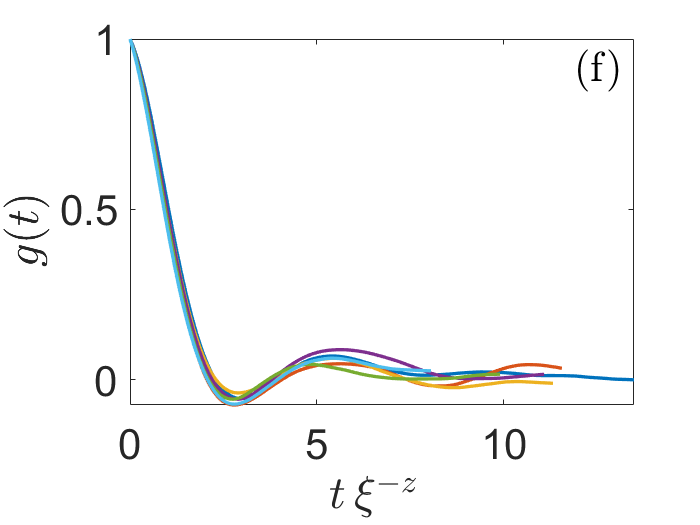}
\end{center}
\caption{{\bf (a)-(c)} Finite-size scaling of correlation length, susceptibility, and polarization as functions of $\beta$ for $\eta=0.5$ and $N=125, 250, 500, 1000, 2000, 4000$ on the critical curve $\beta_0$. {\bf (d)} Determination of the dynamic critical exponent by LS fitting of correlation time versus correlation length. We find $\nu=0.36\pm 0.04$, $\gamma=0.91\pm 0.21$, $b=1.3\pm 0.2$ ($W\sim\beta^b$), $z=1.26\pm 0.10$. {\bf (e), (f)} Partial collapse of the DCF with scaled time $t/\xi^z$ (visually, $z=1.15$) for $\eta=0.5$ and $125\leq N\leq 4000$. } \label{fig4}
\end{figure}
\end{center}

\subsection{Critical exponents of natural swarms}
Like its isotropic counterpart, the anisotropic HCVM exhibits a scale-free chaos phase transition. However, anisotropic confinement modifies the associated critical behavior: the resulting static critical exponents lie between those of the isotropic three-dimensional HCVM \cite{gon24} and its two-dimensional counterpart \cite{gon2D}.

We first fix the noise amplitude at $\eta=0.5$, well inside the chaotic region as in Ref.~\onlinecite{gon23}, and determine the critical exponents from the finite-size dependence of the critical curve $\beta_0(N;\eta)$ for $125\leq N\leq 4000$, spanning the range observed in natural swarms \cite{att14,cav23}. Figs.~\ref{fig4}(a) to \ref{fig4}(c) yield $\nu=0.36\pm 0.04$, $\gamma=0.91\pm 0.21$, $z=1.26\pm 0.10$. Figs.~\ref{fig4}(d) and \ref{fig4}(e) show the partial collapse of the DCF when time is rescaled as $t/\xi^z$ (visually, $z=1.15$), which agrees with that observed in natural swarms \cite{cav17}.

\begin{center}
\begin{figure}[ht]
\begin{center}
\includegraphics[width=0.3\linewidth]{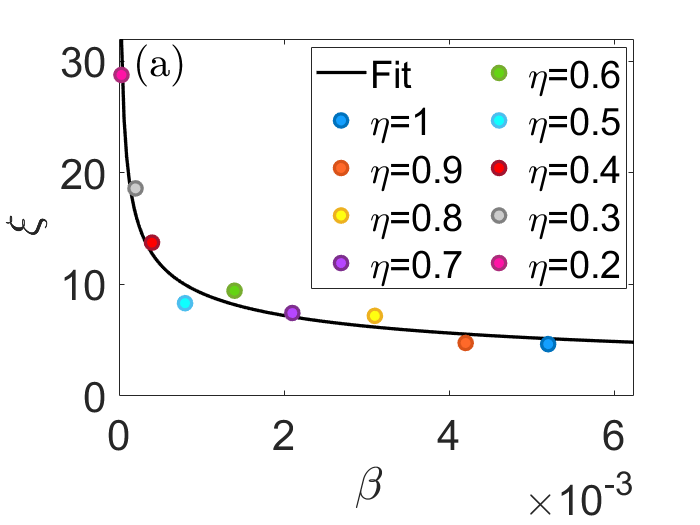}
\includegraphics[width=0.3\linewidth]{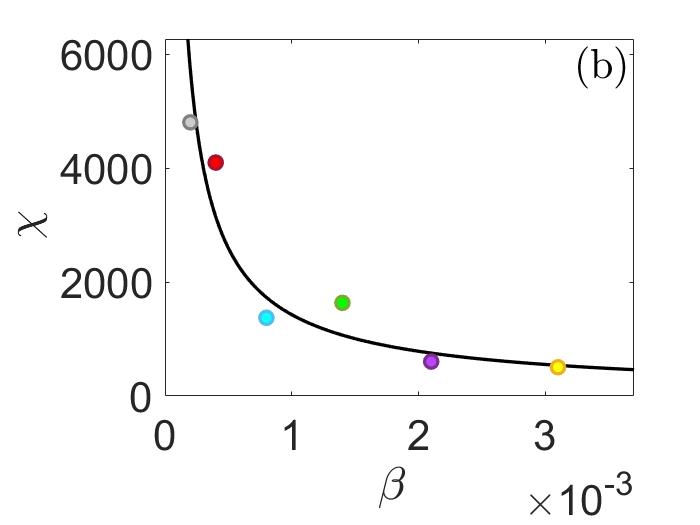}
\includegraphics[clip,width=0.3\linewidth]{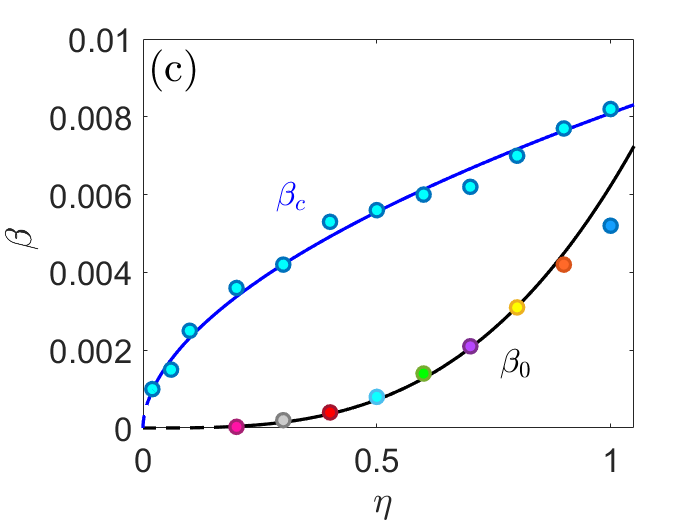}
\includegraphics[clip,width=0.3\linewidth]{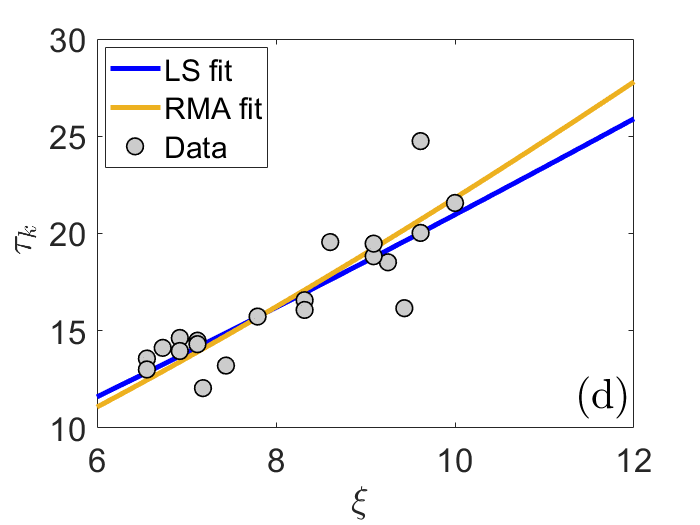}
\includegraphics[width=0.3\linewidth]{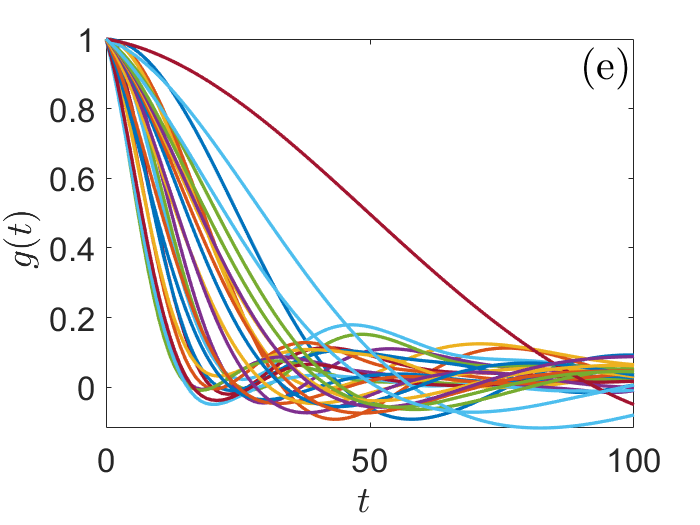}
\includegraphics[width=0.3\linewidth]{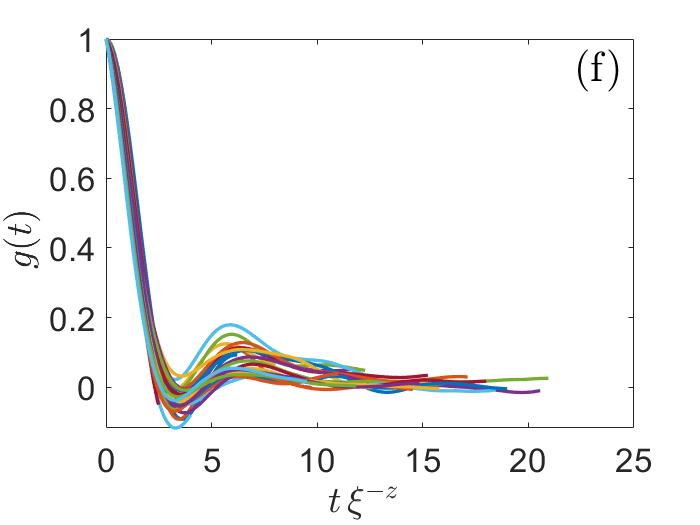}
\end{center}
\caption{ {\bf (a,b)} Finite-size scaling of the correlation length and susceptibility as functions of $\beta$ for different values of the noise $\eta$ on the critical curve $\beta_0(500;\eta)$. {\bf (c)} Extended critical region between the curves $\beta_0$ and $\beta_c$ for $N=500$. The blue curve is the fit $\beta_c = c_c\eta^{m_c}$, with $m_c= 0.54\pm 0.02$ and $c_c= 0.0081 \pm 0.0003$. {\bf (d)} Determination of the dynamic critical exponent from least-squares (LS) and  reduced-major-axis (RMA) fits of the correlation time $\tau_k$ versus the correlation length $\xi$. The resulting estimates are $z_{LS}=1.17 \pm 0.15$, and $z_{RMA}=1.33\pm0.15$. {\bf (e,f)} Partial collapse of the DCF for a combined dataset including simulations on $\beta_0(N;0.5)$ ($125\leq N\leq 4000$), $\beta_0(500;\eta)$ and $\beta_c(500;\eta)$. The best visual collapse is obtained for $z\approx1.1$. } \label{fig5}
\end{figure}
\end{center}

An independent determination of the exponents provides a consistency check on these results. To verify the robustness of these estimates, we also determine the critical exponents by fixing $N$, depicting $\xi$, $\chi$ and $\tau_k$ in terms of $\beta_0$ for different noise values as shown in Fig.~\ref{fig5}(a) and \ref{fig5}(b), and using Eqs.~\eqref{eq13} as explained in Ref.~\cite{gon24}. We obtain $\nu=0.35\pm 0.03$, $\gamma=0.86\pm 0.13$, $z=1.06\pm 0.03$. These values are close to those found by fixing noise and changing $N$. Taking the average of the two independent estimates gives representative values: $\nu = 0.36 \pm 0.04$, $\gamma = 0.89 \pm 0.17$, $z = 1.16 \pm 0.07$, which are compatible with both estimates.

Finally, we investigate the low-noise regime by computing the curve $\beta_c(500;\eta)$ of the anisotropic HCVM for small noise values. The result is shown in Fig.~\ref{fig5}(c). On this curve, we obtain the same critical exponents as before (LS fit) and a new power law for the LLE: $\lambda_1\propto\beta_c^\varphi$, with $\varphi=z\nu$ (the same relation holds for the isotropic HCVM \cite{gon23}). To reproduce the variability observed in natural swarms, we consider a mixture of simulations drawn from the extended critical region bounded by $\beta_0(500;\eta)$ and $\beta_c(500;\eta)$ as follows. We choose points $(\eta,\beta)$ on the lines $\beta_0(500;\eta)$ and $\beta_c(500;\eta)$ in Fig.~\ref{fig5}(d), and numerically simulate Eqs.~\eqref{eq11} with different random initial conditions. The critical exponent $z$ is determined from Eqs.~\eqref{eq6}(b) by fitting the resulting data using LS and RMA, as depicted in Fig.~\ref{fig5}(d). We obtain $z_{LS}=1.17 \pm 0.15$, $z_{RMA}=1.33\pm0.15$, which are close to the values obtained for the isotropic HCVM \cite{gon24}. Figures \ref{fig5}(e) and \ref{fig5}(f) illustrate the partial collapse of the DCF for small values of the scaled time $t\xi^{-z}$. As explained in Ref.~\onlinecite{gon23}, the correlation time as calculated in Eqs.~\eqref{eq6} captures only the longest time scale of the multifractal chaotic attractor. The latter has different time scales as a particle spends different times visiting different parts of the attractor, which accounts for the partial collapse of the DCF.

\begin{table}[ht]
\centering
\begin{tabular}{lccc}
\toprule
System & $\nu$ & $\gamma$ & $z$ \\
\hline
ISM \& RG 
& 0.748 & 1.171 & 1.35\\

effective one-particle
& - & - & 1\\

HCVM (mean field)
&0.5 & 1& 1\\

HCVM (2D, isotropic) 
& $0.30 \pm 0.02$ 
& $0.78 \pm 0.05$ 
& $0.99 \pm 0.03$ \\

HCVM (3D, isotropic) 
& $0.436 \pm 0.009$ 
& $0.92 \pm 0.05$ 
& $1.01 \pm 0.01$ \\

HCVM (3D, anisotropic) 
& $0.36 \pm 0.04$ 
& $0.89 \pm 0.17$ 
& $1.16 \pm 0.07$ \\

Experiments 
& $0.35 \pm 0.10$ 
& $0.9 \pm 0.2$ 
& $1.16 \pm 0.12$ \\
\hline
\end{tabular}
\caption{Comparison of the static ($\nu$, $\gamma$) and dynamic ($z$) critical exponents predicted by the mean-field HCVM, the isotropic HCVM in two and three dimensions, the anisotropic three-dimensional HCVM (without data mixing), the ISM, the effective one-particle theory, and experimental measurements. All exponents are obtained from least-squares (LS) regression fits. The corresponding reduced-major-axis (RMA) estimates of $z$ are given in the text.}
\label{table1}
\end{table}

The critical exponents obtained from mean-field \cite{gon23mf}, 2D \cite{gon2D}, isotropic 3D \cite{gon24}, and anisotropic HCVM (cf Figs.~\ref{fig4} and \ref{fig5}) have been compiled in Table~\ref{table1}, which also includes those measured in natural swarms \cite{att14,cav23} and the other theories described in Section \ref{sec:2}. We observe that the static critical exponents obtained for the anisotropic 3D HCVM lie between the corresponding 2D and isotropic 3D values. Both the static critical exponents, $\nu$ and $\gamma$, agree with the experimental estimates within their uncertainties. This suggests that anisotropic confinement induces an intermediate critical behavior between two and three dimensions, providing a natural mechanism for reconciling theoretical predictions with the experimentally observed static and dynamic critical exponents. In short, anisotropy shifts the HCVM critical exponents toward the experimental values while preserving the scale-free chaos transition. As summarized in Table \ref{table1}, the anisotropic HCVM therefore provides the closest agreement with the experimental critical exponents among the theoretical approaches considered in this review.

\subsection{Effective spring constants and velocity and position densities of natural swarms}
\begin{figure}[ht]
\centering
\includegraphics[width=0.4\textwidth]{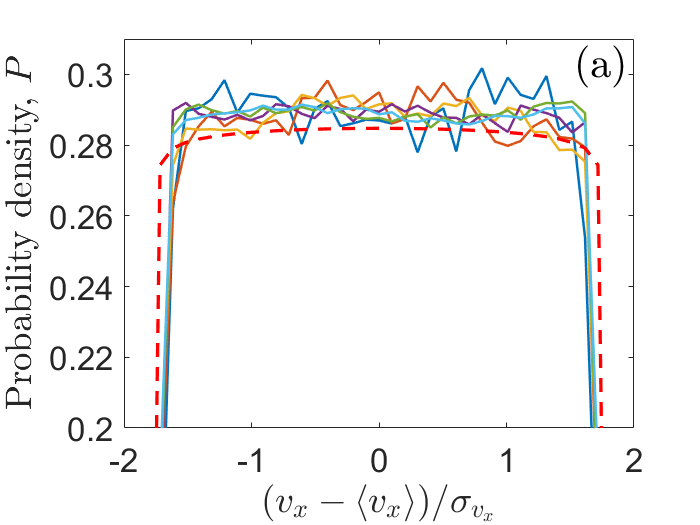}
\includegraphics[width=0.4\textwidth]{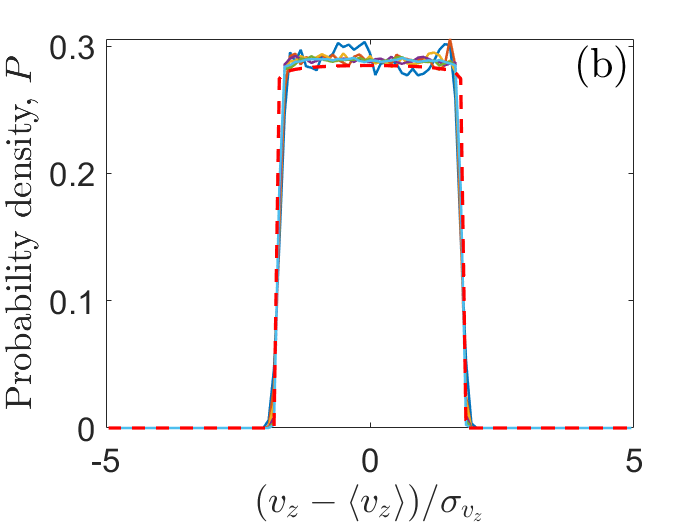}
\includegraphics[width=0.3\textwidth]{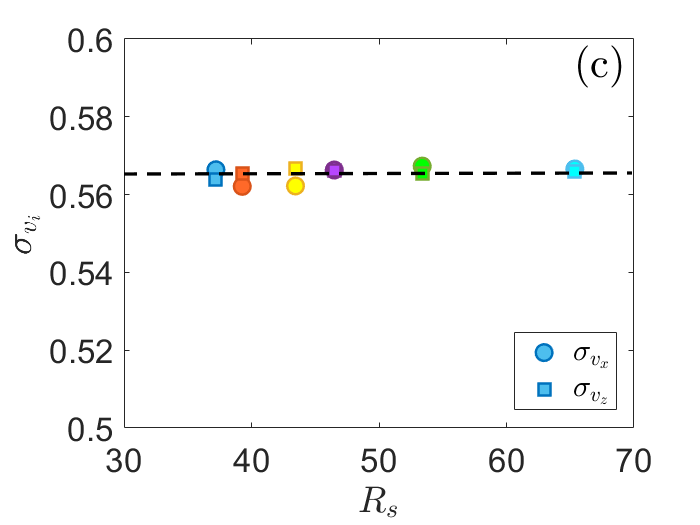}
\includegraphics[width=0.3\textwidth]{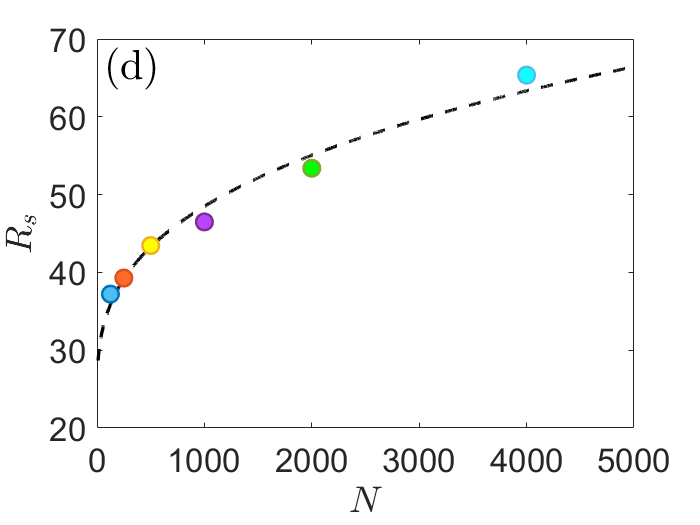}
\includegraphics[width=0.3\textwidth]{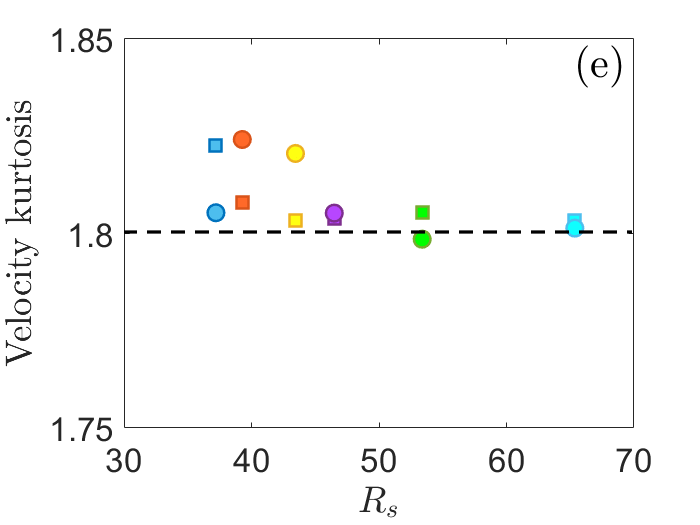}
\caption{ (a,b) Standardized probability density functions (PDFs) of (a) normalized horizontal velocity and (b) vertical velocity compared with a reference $q$-Gaussian distribution ($q=-100<1$) (red discontinuous line). (c) Standard deviation $\sigma_{v_x}$ (circles) and $\sigma_{v_z}$ (squares) versus RMS swarm radius $R_s=\langle \sqrt{x^2+y^2+z^2} \rangle$ (the discontinuous line is $\sigma=0.565$). (d) RMS radius $R_s$ versus $N$ (the dashed line is $R_s=N^{1/3}$). (e) Velocity kurtoses $\langle(\frac{v_x-\langle v_x\rangle}{\sigma_{v_x}})^4\rangle$ and $\langle(\frac{v_z-\langle v_z\rangle}{\sigma_{v_z}})^4\rangle$ versus $R_s$. Colors of the points on the critical line $\beta_0(N;0.5)$ are: $N=125$ (dark blue), 250 (red), 500 (yellow), 1000 (magenta), 2000 (green), 4000 (light blue), respectively.} 
\label{fig6}
\end{figure}

\begin{figure}[ht]
\centering
\includegraphics[width=0.3\textwidth]{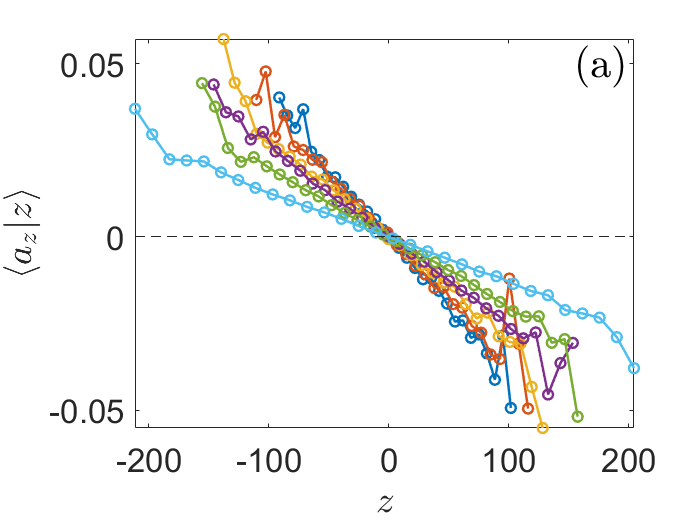}
\includegraphics[width=0.3\textwidth]{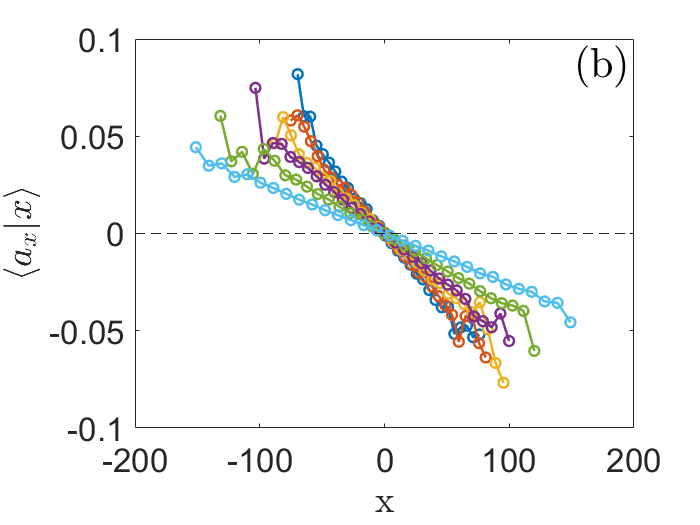}
\includegraphics[width=0.3\textwidth]{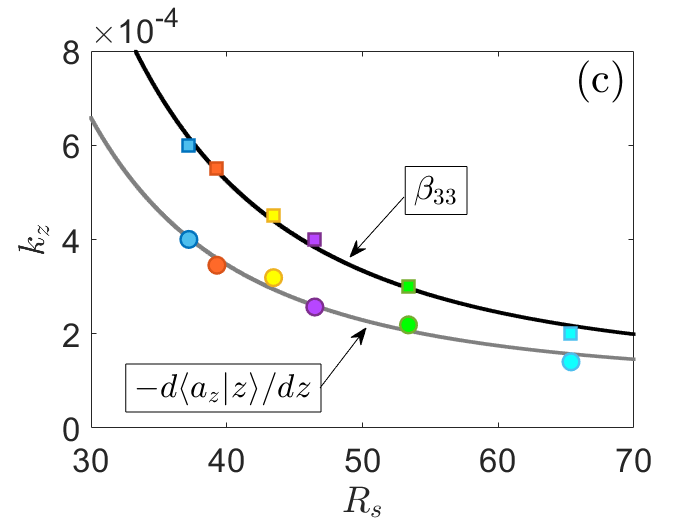}
\includegraphics[width=0.3\textwidth]{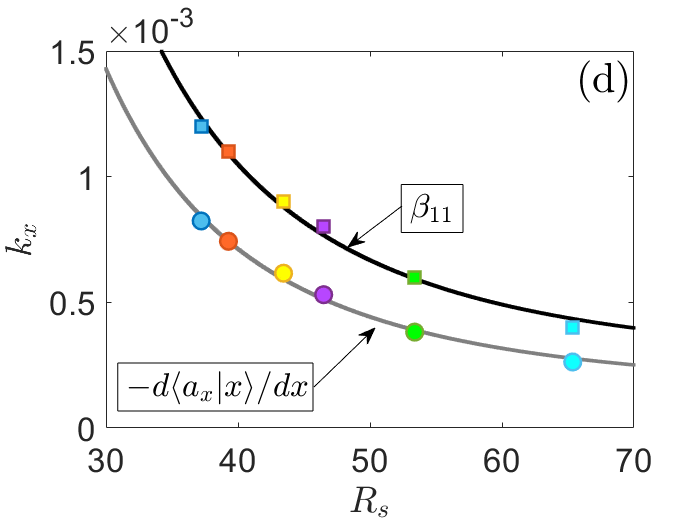}
\includegraphics[width=0.3\textwidth]{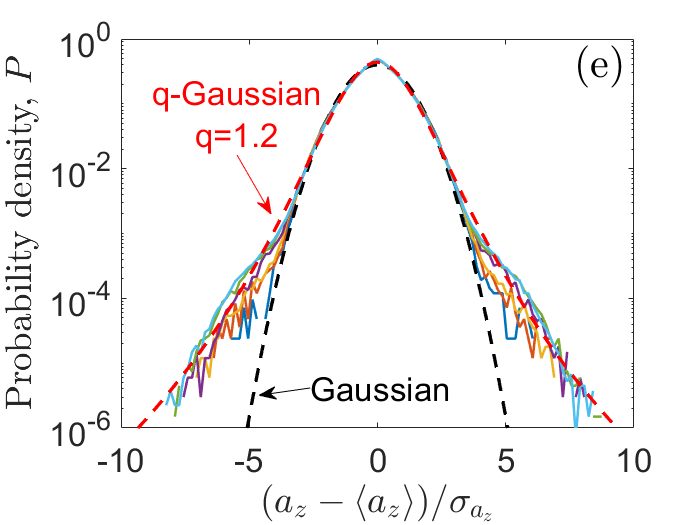}
\includegraphics[width=0.3\textwidth]{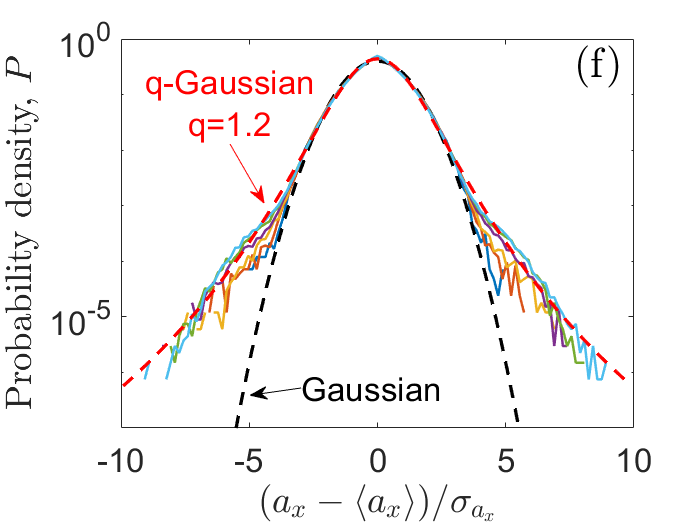}
\caption{ {\bf Mean acceleration as a function of position and PDFs}. (a,b) Mean
acceleration conditioned on position in (a) the vertical direction, and (b)
one horizontal direction. The conditional acceleration varies linearly with
a negative slope as a function of position. (c,d) Effective spring constants
extracted from linear fits to the data in (a) and (b). The effective elastic potentials are stiffer in both directions for smaller swarms. (e,f) Standardized PDFs of vertical (e) and horizontal (f) accelerations exhibit long tails. Colors are as in Fig.~\ref{fig6}, $\beta_{11}=\beta$, $\beta_{33}=\beta/2$.}
\label{fig7}
\end{figure}

\begin{figure}[ht]
\centering
\includegraphics[width=0.3\textwidth]{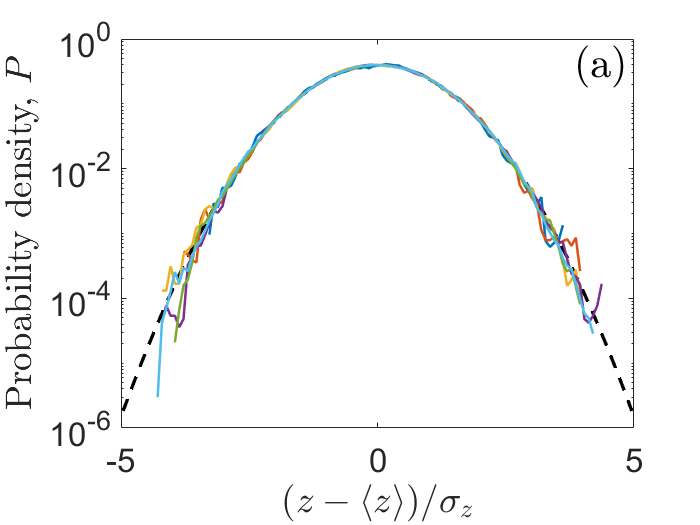}
\includegraphics[width=0.3\textwidth]{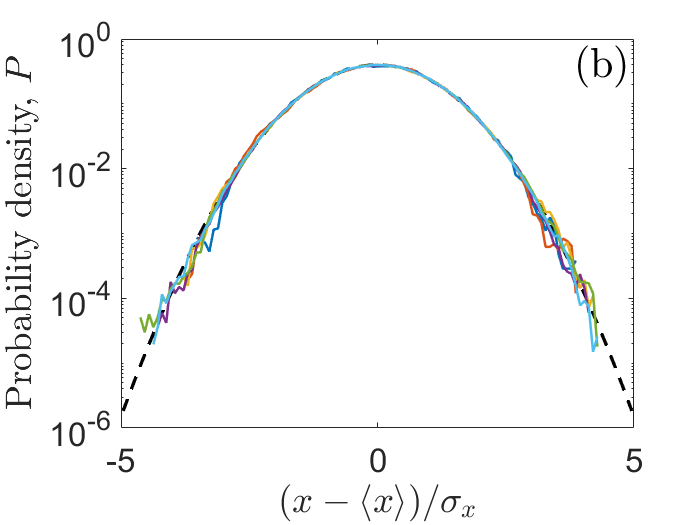}
\includegraphics[width=0.3\textwidth]{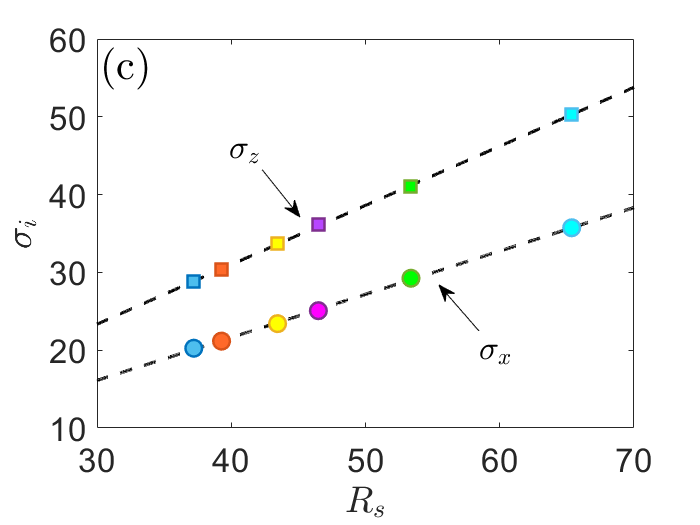}
\includegraphics[width=0.3\textwidth]{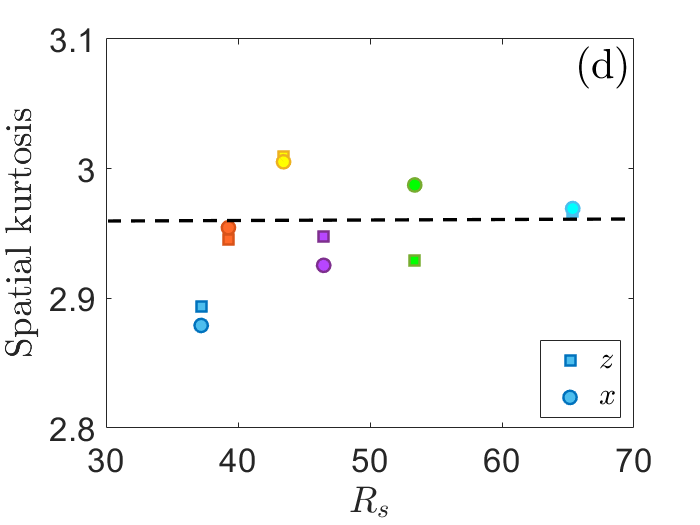}
\caption{{\bf PDF as a function of position.}
(a,b) Time and velocity averaged PDF of (a) relative vertical direction and (b) relative horizontal direction. The discontinuous lines are Gaussian PDFs. (c) Standard deviations $\sigma_z$ (squares) and $\sigma_x$ (circles) in the vertical and horizontal directions, respectively, versus RMS radius. We have $\sigma_{x}\sim 0.5 R_s$, $\sigma_{z}\sim 0.82 R_s$. (d) The respective kurtoses are close to 3 (Gaussian PDF). }
\label{fig8}
\end{figure}

We now investigate the long-time probability density functions (PDFs) generated by the anisotropic HCVM and compare them with the available experimental observations. The PDFs depend on the location of the system in the phase diagram. Unless otherwise stated, all quantities are computed after discarding an initial transient period. Means and standard deviations are obtained by averaging over particles and time, while the PDFs are constructed from the fraction of time that particles spend within prescribed intervals of the corresponding standardized variables. On the critical line $\beta_0(N;\eta)$ with 
$\eta=0.5$, and $125\leq N\leq 4000$, we calculate effective spring constants and the marginal PDFs shown in Figures \ref{fig6} - \ref{fig8}.

Because the equations of motion are integrated in discrete time, the velocities and accelerations used in the analysis require some discussion. Whereas experiments sample the underlying continuous trajectories of insects \cite{kel13,att14plos,att14}, the HCVM generates a discrete-time dynamical system. In both cases, however, velocities and accelerations are reconstructed from discretely sampled positions by numerical smoothing and differentiation. Thus, the same analysis pipeline is applied to both the experimental and simulated datasets.

Time derivatives are calculated using convolution with a Gaussian smoothing-and-differentiation kernel, as described in \cite{mor04}. Consequently, the reconstructed velocities and accelerations, as well as derived quantities such as their probability density functions and effective spring constants, depend quantitatively on the sampling interval and the smoothing width. Throughout this work, these parameters are kept fixed, and the same reconstruction protocol is applied to all simulations being compared.

The Gaussian kernel used throughout the analysis has width $\sigma=w\Delta t$ ($w=2$). We have verified that moderate variations of the kernel width do not affect the results.

Figure~\ref{fig6} summarizes the statistics of the vertical and horizontal velocities along the critical line $\beta_0(N;\eta=0.5)$. Unlike the nearly Gaussian velocity PDFs measured in laboratory swarms \cite{kel13}, the simulated distributions are well approximated by compact-support $q$-Gaussians ($q<1$; the Gaussian corresponds to 
$q=1$) \cite{tsa23},
\begin{eqnarray}
f_q(x)= \frac{1}{C_q} [1-(1-q) x^2]_+^\frac{1}{1-q}, \quad C_q=\int_{-\infty}^\infty [1-(1-q) x^2]_+^\frac{1}{1-q} dx,\quad [x]_+=x\,\theta(x),  \label{eq15}
\end{eqnarray}
defined in Eq.~\eqref{eq15}. As shown in Figs. 6(a,b), the numerical PDFs are well described by this functional form. Their standard deviations remain close to 0.565 [Fig.~\ref{fig6}(c)], while the swarm radius scales as $R_s\sim N^{1/3}$ [Fig. ~\ref{fig6}(d)]. The polarization decreases with increasing swarm size (not shown), the speed statistics is radially independent (not shown), and the kurtosis approaches 1.8, significantly below the Gaussian value of 3 [Fig.~\ref{fig6}(e)].
 
Following the experimental analysis of Ref.~\cite{kel13}, we next examine how the acceleration depends on the position within the swarm. Fig.~\ref{fig7}(a) and \ref{fig7}(b) show the mean vertical acceleration conditioned on vertical and horizontal position, respectively. Similarly to laboratory swarms, midges above (below) the  swarm center accelerate downwards (upwards). A similar trend of acceleration towards the swarm center is clear in the horizontal direction. The approximately linear dependence of the conditional acceleration on position is consistent with an effective harmonic restoring force toward the swarm center. Consistent with the confinement parameters  $\beta_{11}=\beta_0$ and $\beta_{33}=\beta_0/2$, the effective spring constants decrease as the RMS swarm radius (and therefore the correlation length) increases [Figs.~\ref{fig7}(c,d)]. Figs.~\ref{fig7}(e) and \ref{fig7}(f) illustrate the PDFs  as functions of vertical and horizontal accelerations, respectively. They can be approximated by $q$-Gaussians with $q>1$ that are wider than Gaussians. These functional shapes are consistent with the interpretation of power laws associated to the scale-free chaos phase transition \cite{gon23,gon24}. Note that similar trends for laboratory swarms (Fig.~6 in \cite{kel13}) cannot be associated to the same phase transition because the correlation length remains small as the insect number increases. 

Finally, the position PDFs along the critical line $\beta_0(N;\eta)$ are close to Gaussian distributions in both the vertical and horizontal directions, as shown in Figs.~\ref{fig8}(a,b).
On the critical line $\beta_0(N;\eta)$, the PDFs of position in the vertical and horizontal directions are close to Gaussians, as demonstrated by Figs.~\ref{fig8}(a) and \ref{fig8}(b). Fig.~\ref{fig8}(c) shows that the standard deviations in both directions are proportional to the swarm RMS radius, and therefore they also serve as proxies of the correlation length. Fig.~\ref{fig8}(d) depicts that the kurtosis is close to 3, thereby corroborating that the PDFs are close to Gaussian functions.

\subsection{Effective spring constants and PDFs of laboratory swarms}
Figures \ref{fig6} - \ref{fig8} show that the statistics predicted for natural swarms differ substantially from those measured in laboratory swarms. This raises the question of whether the latter can be reproduced by exploring a different region of the HCVM phase diagram. We restrict attention to laboratory observations that do not require controlled external forcing, namely PDFs close to Gaussian distributions, the layered (``slab'') structure of the position PDFs, and the dependence of the effective spring constants on swarm size for $N<100$. We investigate two complementary strategies: (i) zero interaction radius, corresponding to weak correlations, and (ii) finite interaction radius within the extended critical region of multicluster chaos.

\subsubsection{Zero interaction radius and weak correlations} 
Laboratory swarms exhibit small correlation lengths and low polarization \cite{kel13}, while the distances between neighboring insects appear approximately scale free even in small swarms \cite{rey23}. Within the HCVM, these observations motivate setting the interaction radius to zero ($R_0=0$), which eliminates spatial correlations. A small polarization can be imposed by randomly placing particles (midges) in points over a chaotic or quasiperiodic attractor of the mean field HCVM corresponding to $N=1$ \cite{gon23mf}, see \cite{bon25} for examples of exact solutions of the deterministic HCVM where small polarization follows from random placing over simpler attractors. Lastly, we need a phase region $(\eta,\beta)$ where the PDF are similar to laboratory data \cite{kel13}.

\begin{figure}[ht]
\centering
\includegraphics[width=0.3\textwidth]{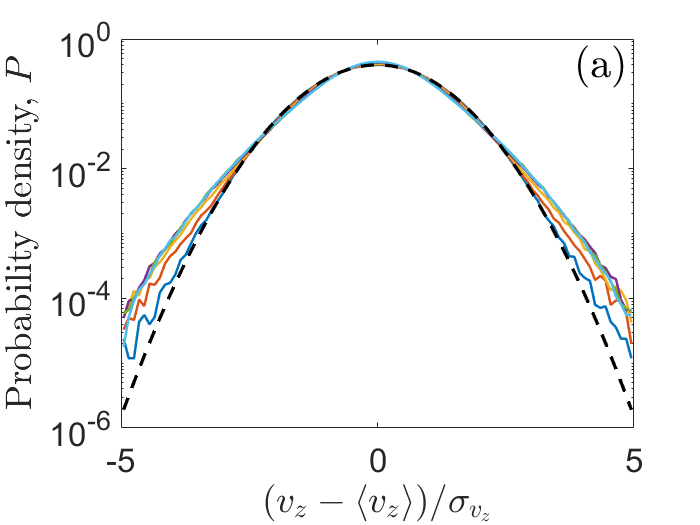}
\includegraphics[width=0.3\textwidth]{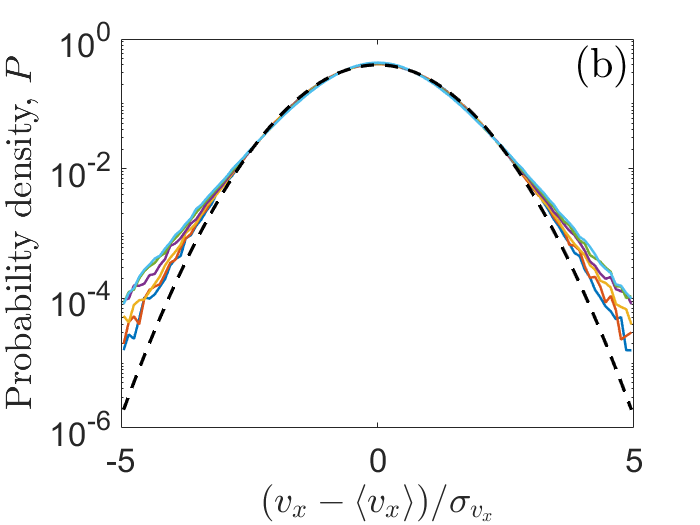}
\includegraphics[width=0.3\textwidth]{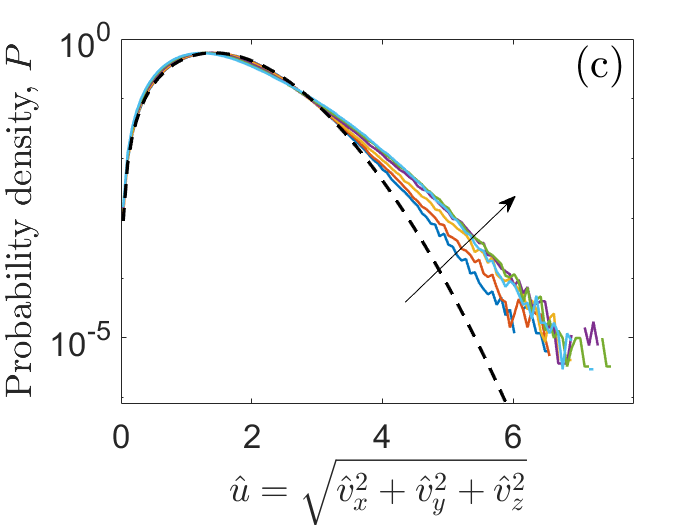}
\includegraphics[width=0.3\textwidth]{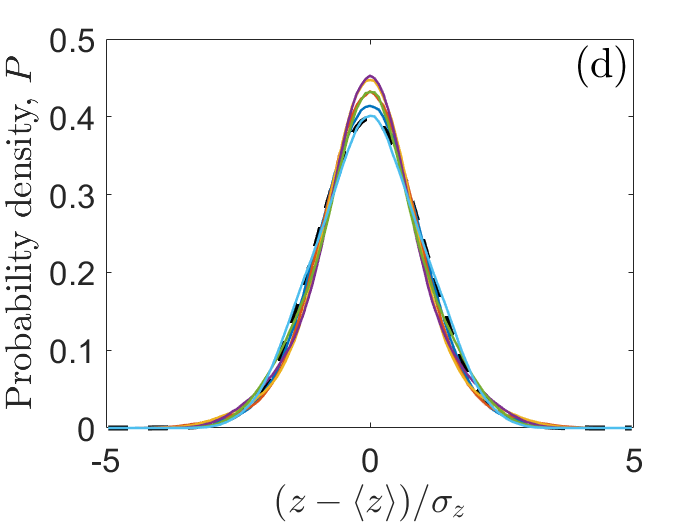}
\includegraphics[width=0.3\textwidth]{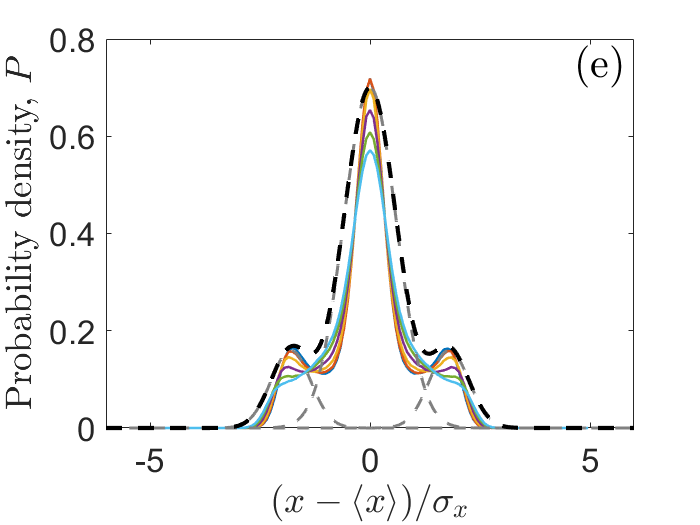}
\includegraphics[width=0.3\textwidth]{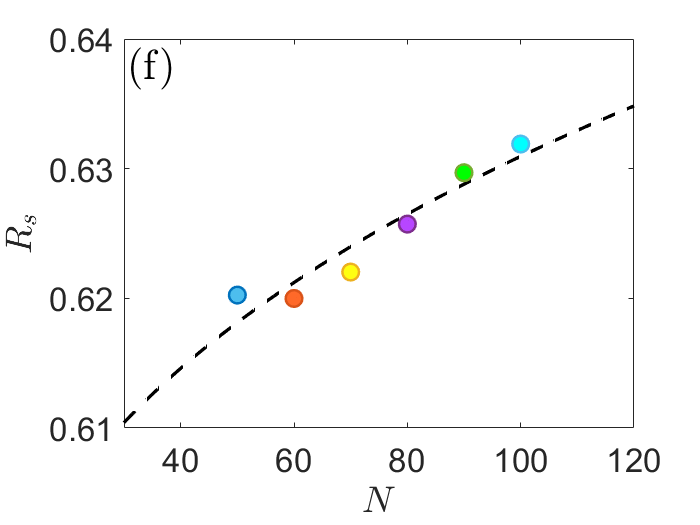}
\caption{{\bf PDF versus velocity and position for $R_0=0$.}
(a,b) Standardized PDFs of (a) normalized vertical velocity and (b) horizontal velocity compared with a reference Gaussian distribution. (c) Standardized PDF of the normalized speed $\hat{u}=\sqrt{\hat{v}_x^2+\hat{v}_y^2+\hat{v}_z^2}$, where $\hat{v}_i=(v_i-\langle v_i\rangle)/\sigma_{v_i}$, compared with the standardized Maxwell-Boltzmann distribution (discontinuous line). A heavy, nearly exponential tail develops for large swarms, which may indicate the formation of clusters. (d,e) Standardized PDFs of (d) normalized vertical position and (e) normalized horizontal position. While the PDF of the vertical position is close to a Gaussian function, the PDF of the horizontal position is close to a superposition of three Gaussian functions. (f) RMS swarm radius $R_s$ versus $N^{1/3}$. Swarms have colors $N=50$ (dark blue), 60 (red), 70 (yellow), 80 (magenta), 90 (green), and 100 (light blue).}
\label{fig9}
\end{figure}

We know that standardized PDFs of the velocities are almost Gaussian \cite{kel13}, which means their kurtoses are about 3. We explore the phase diagram for a fixed $\eta=0.5$ \cite{gon23} and $50<N<100$ by selecting $\beta$ so that the velocity kurtoses are about 3. For a range $50<N<100$, this produces a confinement interval $2.7<\beta<3.2$ within the region of noisy chaos but far from the critical lines corresponding to the scale-free chaos phase transition.  Then we try the values $(\beta,N)= (2.7,50)$, $(2.8,60)$, $(2.9,70)$, $(3.0,80)$, $(3.1,90)$, $(3.2,100)$. An influence sphere of zero radius enforces zero correlation length for the swarm. To ensure a low polarization, the particles are randomly distributed within the unit sphere at the initial time and they become randomly distributed on the attractor after a transient period.  The resulting polarization is below 0.13 and decreases as $R_s$ increases. Figs.~\ref{fig9}(a) to \ref{fig9}(c) show that the standardized PDF of the velocities and the speed develop heavy tails and deviate from the Gaussian for increasing $N$ (indicated by the arrow in Fig.~\ref{fig9}(c)). These trends are similar to those found in experiments \cite{kel13}. However, the standardized PDF of the vertical positions are close to a Gaussian function, cf Fig.~\ref{fig9}(d), whereas the PDF of the horizontal positions are a superposition of three Gaussian functions, as seen in Fig.~\ref{fig9}(e). This indicates that the horizontal cross-sectional area of the  ellipsoid-shaped swarms increases with the population size while their height remains constant. Although this behavior has not yet been reported experimentally, it is biologically plausible because it is advantageous for swarming males to be flying at the typical height of incoming females, and because flying upwards (against gravity) to explore the swarm volume in its entity is energetically costly. In data for small swarms, the situation is just the opposite: the PDFs of vertical positions are a superposition of Gaussian functions (`slabs') whereas the PDFs of horizontal positions are single Gaussian functions, cf Fig.~1 of \cite{rey21}. Thus, the simulated velocity PDFs are compatible with lab observations but the simulated position PDFs are not. Additionally, Fig.~\ref{fig9}(f) shows that the swarm RMS radius scales as $N^{1/3}$ for $50<N<100$.

We also explored modifying the anisotropy by setting $\beta_{33}=2\beta$. This recovers the layered structure in the vertical direction and yields $k_z<k_x$, but both effective spring constants then increase with $N$, worsening the agreement with experiment. Alternative choices in which $\beta$ decreases with $N$ improve some observables while degrading others. Thus, the mild increase of $\beta$ from 2.7 to 3.2 provides the best overall compromise for reproducing the observed velocity, speed, and acceleration statistics.

\begin{figure}[ht]
\centering
\includegraphics[width=0.3\textwidth]{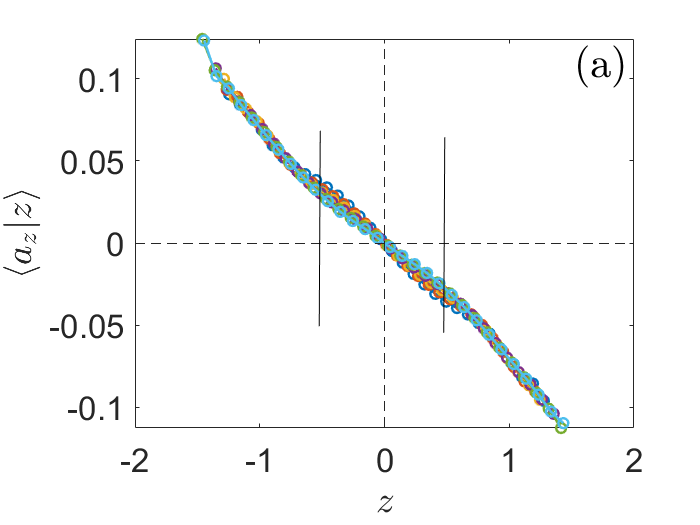}
\includegraphics[width=0.3\textwidth]{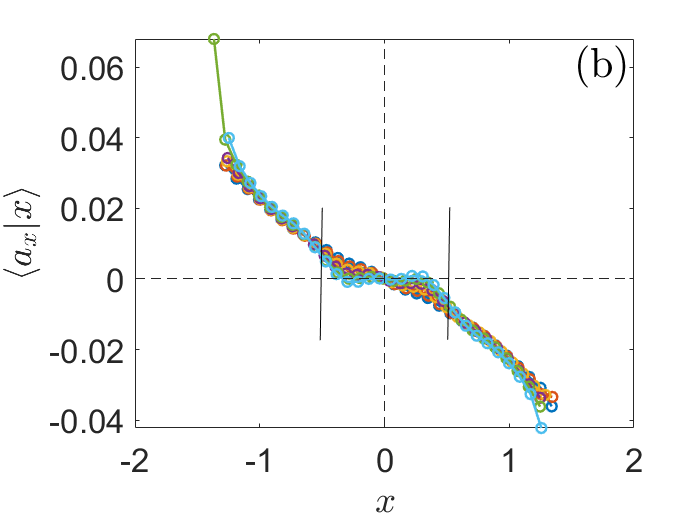}
\includegraphics[width=0.3\textwidth]{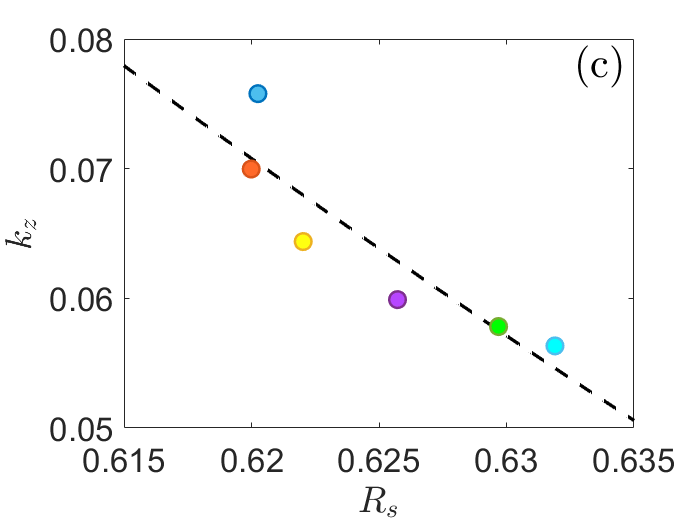}
\includegraphics[width=0.3\textwidth]{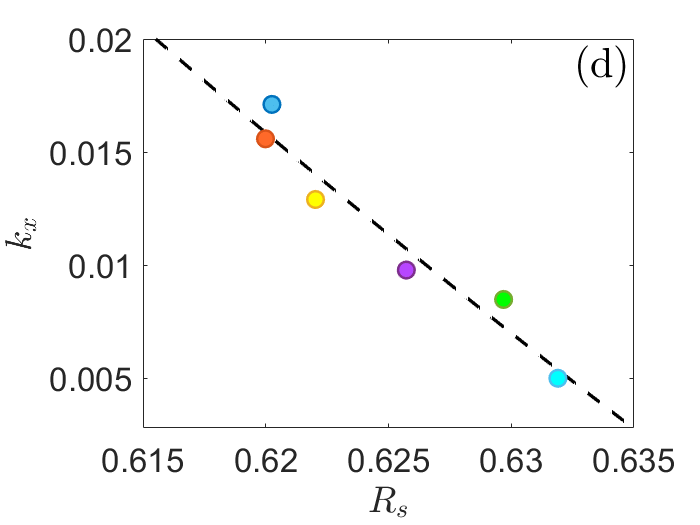}
\includegraphics[width=0.3\textwidth]{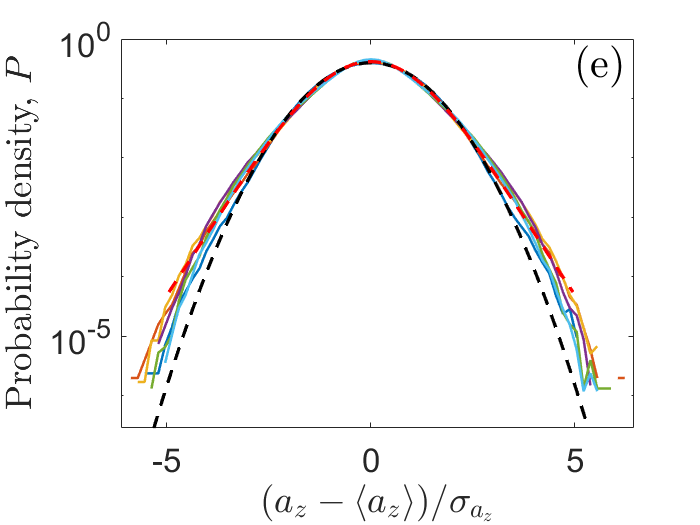}
\includegraphics[width=0.3\textwidth]{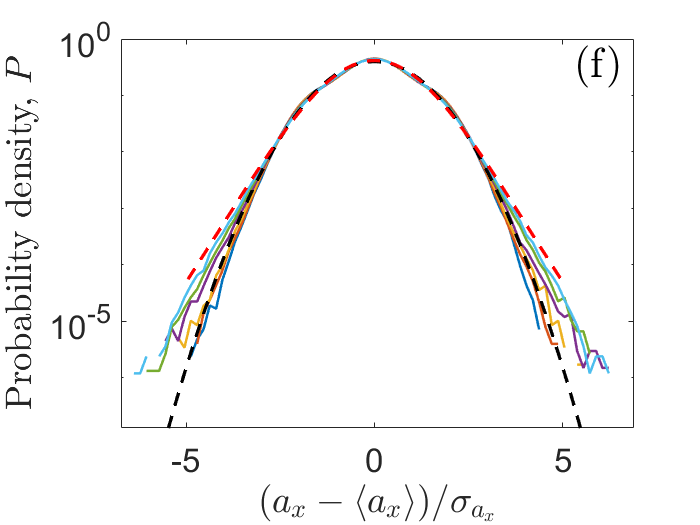}
\caption{{\bf Mean acceleration as a function of position and PDFs for $R_0=0$}. (a,b) Mean acceleration conditioned on position in (a) the vertical direction, and (b) one horizontal direction. The conditional acceleration varies linearly with
a negative slope as a function of position. (c,d) Effective spring constants
extracted from linear fits to the data in (a) and (b). The effective elastic potentials are stiffer in both directions for smaller swarms. (e,f) Standardized PDFs of vertical (e) and horizontal (f) accelerations exhibit long tails. They are better fitted by $q$-Gaussian functions with $q=1.1$ (dashed red lines) than by Gaussians ($q=1$, dashed black lines). Colors are as in Fig.~\ref{fig9} and $(\beta,N)= (2.7,50)$, $(2.8,60)$, $(2.9,70)$, $(3.0,80)$, $(3.1,90)$, $(3.2,100)$.}
\label{fig10}
\end{figure}

What about effective spring constants and acceleration statistics? Figs.~\ref{fig10} are quite similar to Figs.~\ref{fig7} for critical confinement $\beta_0(N;\eta)$ which, in turn, are similar to Figures 5 and 6 of \cite{kel13}. However, the vertical accelerations and spring constants are larger than the corresponding horizontal magnitudes, cf. Fig.~\ref{fig10}(a) to \ref{fig10}(d), which is the opposite to Figs.~\ref{fig7} (critical confinement) and to experimental data \cite{kel13}. The acceleration PDFs are closer to Gaussians in Figs.~\ref{fig10}(e) and \ref{fig10}(f) than the corresponding PDFs of Fig.~\ref{fig7} and of the experimental data \cite{kel13}. Overall, eliminating interactions reproduces the weakly correlated nature of laboratory swarms but fails to capture their observed spatial organization.

\subsubsection{Finite interaction radius within the critical region of multicluster chaos}
The second strategy retains a finite interaction radius and explores the extended critical region associated with multicluster chaos. For small $N$ and finite interaction radius, the critical lines of the scale-free chaos transition within the extended criticality region separate from one another, whereas they all converge to zero confinement as $N\to\infty$ \cite{gon23,gon24}. Then points farther from $\beta_0(\eta;M)$ within the extended criticality region with positive LLE could exhibit PDFs closer to Gaussians as in the experiments \cite{kel13}.  We select the parameters $R_0=5$, $ \eta=0.1$, and 
\begin{subequations} \label{eq16}
\begin{eqnarray}
    (\beta,N)=
    (0.0056,50),(0.0061,60),(0.0067,70),
    (0.0073,80),(0.0079,90),(0.0085,100).\quad
    \label{eq16a}
\end{eqnarray}
Although $\beta$ increases slightly with $N$, the ratio $\beta/N$ decreases:
\begin{equation}
    \frac{\beta}{N}
    =
    1.120,1.017,0.957,0.912,0.878,0.850
    \times 10^{-4},
    \label{eq16b}
\end{equation}\end{subequations}
and the swarm radius increases with $N$ even though the confinement parameter increases. The simulated points are located between $\beta_i$ and $\beta_m$, where $\beta_i$ corresponds to the maximum of the LLE as a function of $\beta$ or, equivalently, to the inflection region of the susceptibility curve, and $\beta_m$ gives the local maximum of the susceptibility for fixed $\eta$ \cite{gon23}. Thus, our parameters lie inside the chaotic multicluster region.  

\begin{figure}[ht]
\centering
\includegraphics[width=0.3\textwidth]{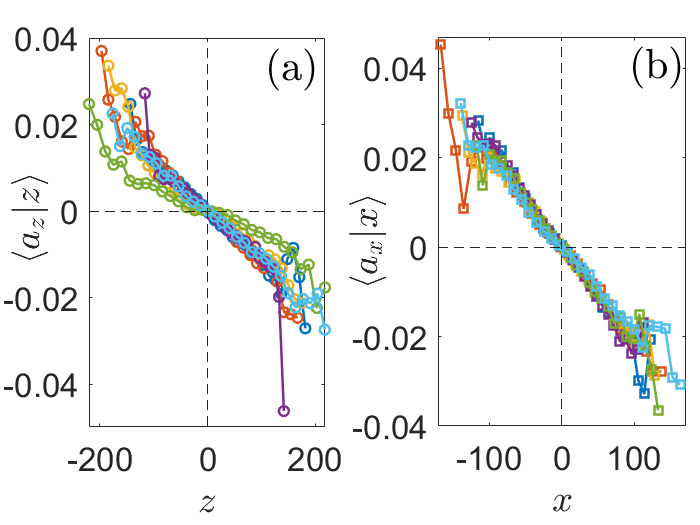}
\includegraphics[width=0.3\textwidth]{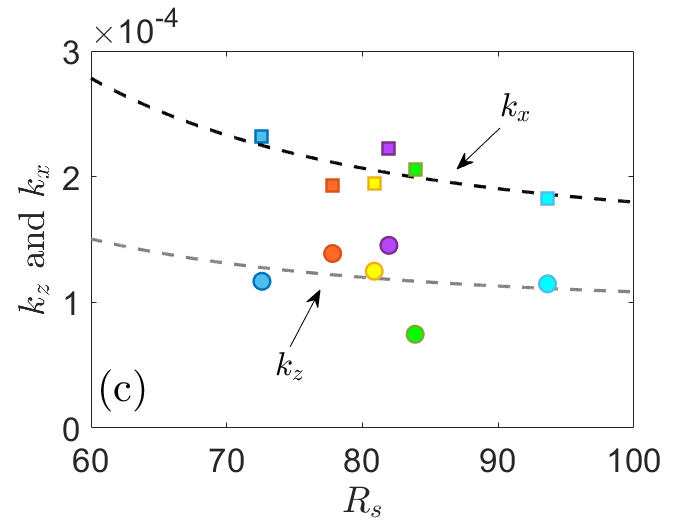}
\includegraphics[width=0.3\textwidth]{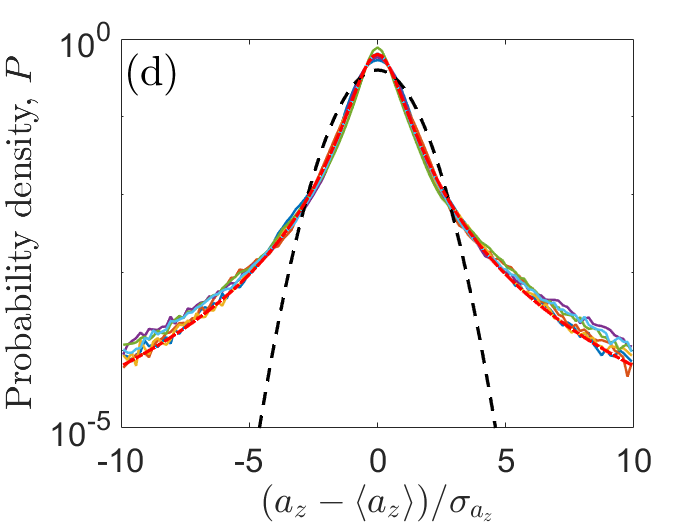}
\includegraphics[width=0.3\textwidth]{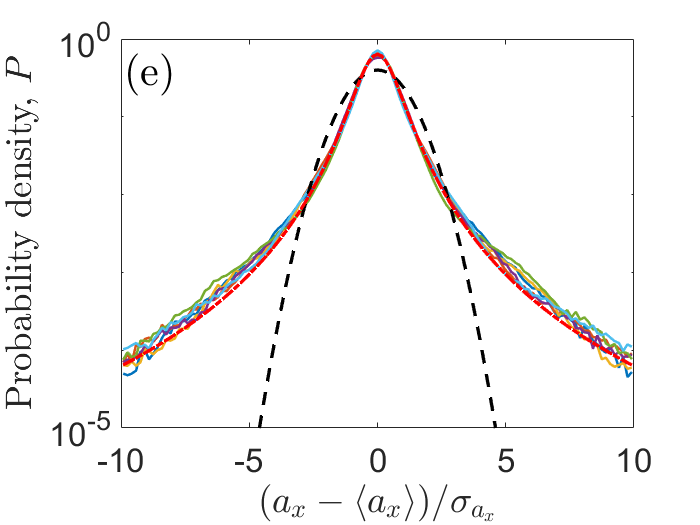}
\includegraphics[width=0.3\textwidth]{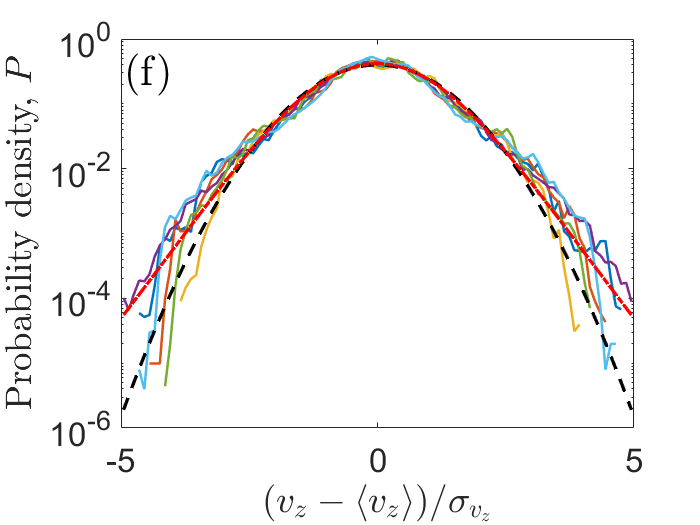}
\includegraphics[width=0.3\textwidth]{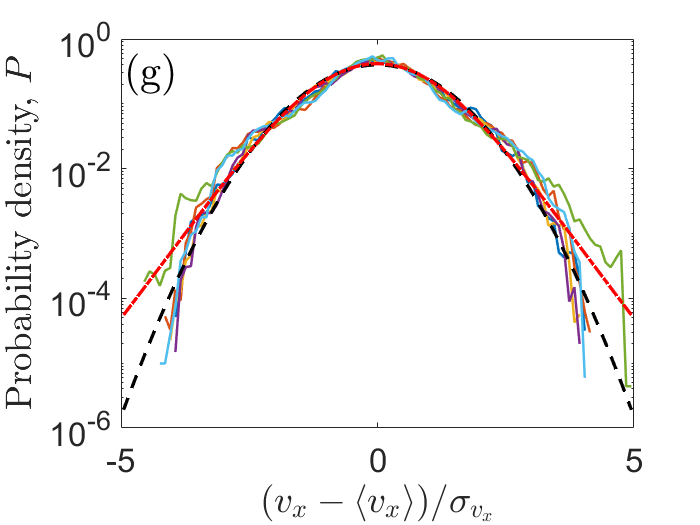}
\includegraphics[width=0.3\textwidth]{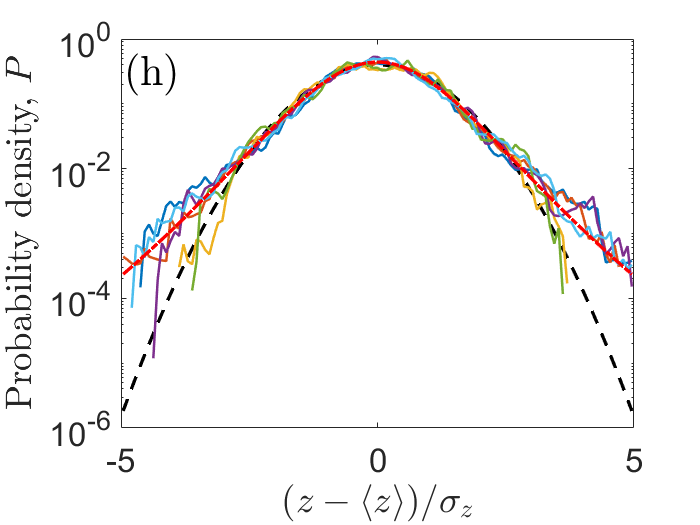}
\includegraphics[width=0.3\textwidth]{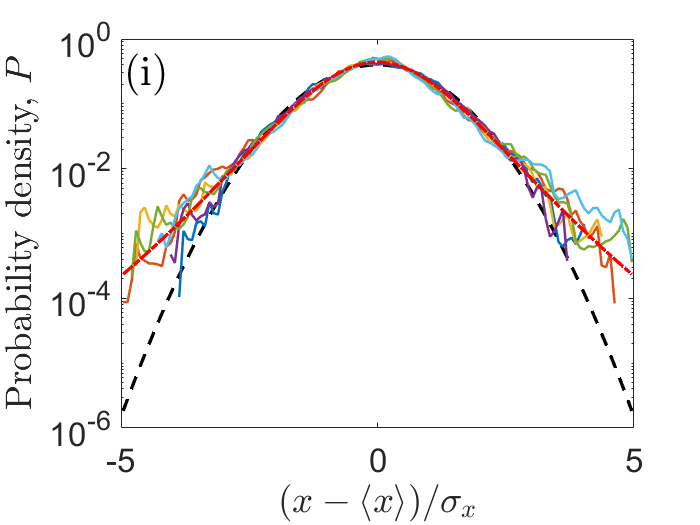}
\includegraphics[width=0.3\textwidth]{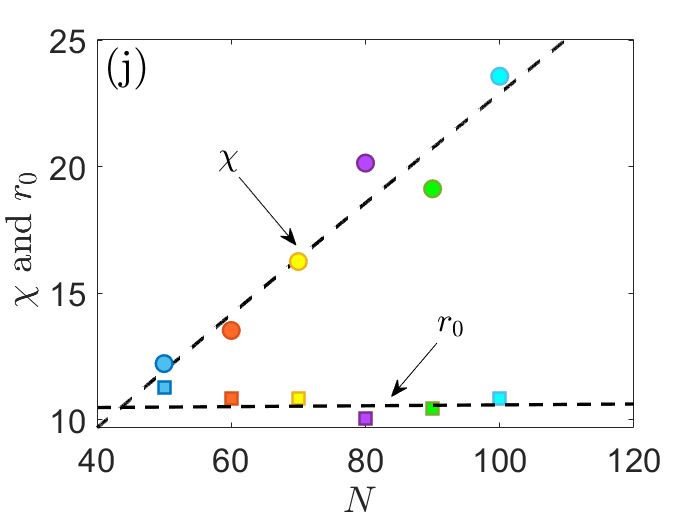}
\caption{
{\bf Mean acceleration as a function of position and PDFs for $R_0>0$}. (a,b) Mean acceleration conditioned on position in (a) the vertical direction, and (b) one horizontal direction. (c) Effective spring constants along the vertical and horizontal directions versus RMS radius. (d,e) Standardized PDFs of vertical (d) and horizontal (e) accelerations exhibit long tails. They are better fitted by $q$-Gaussian functions with $q=1.5$ (dashed red lines) than by Gaussians ($q=1$, dashed black lines). (f,g) Standardized PDFs of (f) normalized vertical velocity and (g) horizontal velocity compared with $q$-Gaussian functions ($q=1.1$, dashed red lines) and to reference Gaussian distributions (dashed black lines). (h,i) Standardized PDFs of vertical (h) and horizontal (i) positions ($q=1.2$-Gaussians, dashed red lines, Gaussians, dashed black lines). (j) Susceptibility and correlation length vs $N$. Colors are as in Fig.~\ref{fig9}.}
\label{fig11}
\end{figure}

Figs.~\ref{fig11} depict the mean accelerations conditioned on position, Panels (a) and (b), the standardized acceleration PDFs, Panel (c), the effective spring constants, Panels (d) and (e), the velocity PDFs, Panels (f) and (g), and the position PDFs, Panels (h) and (i). The effective spring constants extracted from the linear part of the conditional accelerations decrease with the RMS radius $R_s$ and satisfy
\begin{eqnarray}
    k_z = 4.83\, R_s^{-1/\nu} + 0.00009,\quad
    k_x = 11.3\, R_s^{-1/\nu} + 0.00015,
    \label{eq17}
\end{eqnarray}
with $\nu=0.36$, cf Fig.~\ref{fig11}(c).  As in the experiments \cite{kel13}, Fig.~\ref{fig11}(j) shows that the correlation length $r_0$ of Eq.~\eqref{eq4} remains approximately constant as $N$ increases, whereas the susceptibility grows as
\begin{subequations}\label{eq18}
\begin{equation}
    \chi \sim N^\alpha,
    \qquad
    \alpha = 0.9406\pm0.0022.
    \label{eq18a}
\end{equation}
Using the finite-size scaling relation
\begin{equation}
    \alpha=\frac{\gamma}{3\nu},
    \label{eq18b}
\end{equation}
we get
\begin{equation}
    \gamma \simeq 3\nu\alpha \simeq 1.02
    \quad \text{for } \nu\simeq0.36,
    \label{eq18c}
\end{equation}\end{subequations}
and the critical exponent of the susceptibility is $\gamma\simeq1$. The PDFs in Figs.~\ref{fig11} also compare favorably with experiments: position, velocity and acceleration PDFs are respectively well represented by reference $q$-Gaussians with
\begin{equation}
    q_x\simeq1.2,\quad q_v\simeq1.1,\quad q_a\simeq1.5.
    \label{eq19}
\end{equation}
However, the polarization is too large, about 0.6 for different values of $N$ if the velocities are calculated from Eqs.~\eqref{eq11}. When the velocities are obtained from numerical differentiation of the positions supplied by the model, the polarization becomes about 0.3, which is close to the measured values $\sim 0.2$. Three simple mo\-difications reduce the polarization: decreasing $R_0$, increasing $\eta$, or decreasing $\beta$. However, each modification worsens comparison of another observable with experimental values. Decreasing $R_0$ or increasing $\eta$ reduces the acceleration kurtosis, making the acceleration PDFs too close to Gaussian. Decreasing $\beta$ keeps the acceleration PDFs closer to the desired form, but the velocity PDFs become more compact and resemble top-hat distributions ($q<1$-Gaussians). Reducing the polarization without simultaneously degrading the velocity or acceleration statistics therefore appears difficult within the present model. Overall, finite interactions reproduce the non-Gaussian statistics and scaling trends but at the cost of a polarization larger than observed. Furthermore, the speeds increase for small radius and reach a plateau for larger radii (not shown).

In conclusion, strategy 1 successfully reproduces the small polarization and the weakly correlated character of laboratory swarms (having small $N$), and gives reasonable velocity and acceleration PDFs. Its main problems are the inverted spatial structure and the ordering of the spring constants. However, this spatial structure could be biologically favorable for larger swarms. Strategy 2 provides a more coherent picture within the region of extended criticality and multicluster chaos: the effective spring constants decrease with $R_s$, the susceptibility grows with an exponent compatible with $\gamma\simeq1$, and the acceleration PDFs are strongly non-Gaussian with long tails. Its main limitation is a larger polarization. Therefore, the two strategies reveal a clear compromise: the low-polarization laboratory regime is easier to reproduce with $R_0=0$, whereas the non-Gaussian statistics and susceptibility growth with appropriate critical exponent are better reproduced inside the finite-$R_0$ multicluster chaotic region. A systematic exploration of the four-dimensional parameter space $(R_0,\eta,\beta_{11},\beta_{33})$ may reveal regions that simultaneously reproduce the full set of laboratory observations.

\section{Discussion}\label{sec:5}
Midge swarms exhibit several puzzling features. In the wild they display strong correlations and scale-free behavior, with a correlation length $\xi$ proportional to the swarm size \cite{att14plos,att14}, despite their small polarization. Quantities derived from correlation functions exhibit power-law behavior with unusual static and dynamical critical exponents \cite{att14,cav17,cav23}. Moreover, the dynamic correlation function collapses only at short rescaled times, while remaining flat as $t/\xi^z\to 0$. In laboratory conditions without external perturbations, midges remain tightly bound to the swarm but exhibit only weak internal correlations and a small correlation length. External perturbations appear to restore scale-free behavior \cite{vaa20}. However, the available laboratory swarms ($5<N<90$; cf. Fig. 2 of  \cite{vaa20}) are too small to establish critical scaling in macroscopic observables \cite{stu12}). Among the theoretical approaches reviewed here, the isotropic and anisotropic HCVM currently account for the broadest range of observations on natural swarms. However, direct experimental evidence for the underlying scale-free chaos transition is still lacking.

Laboratory swarms exhibit a different set of characteristic properties. They have effective spring constants that decrease with RMS radius, susceptibility with static critical exponent close to 1, non-Gaussian acceleration, position and velocity PDFs with long tails. These and other observations have been successfully described by effective one-particle theories. However, these theories do not include explicit microscopic interactions between midges or the resulting spatial correlations and therefore do not presently account for the critical behavior observed in natural swarms. Reconciling these apparently disparate observations within a single theoretical framework remains a major challenge. The evidence reviewed in this article suggests that significant progress has been made, but important questions remain open.

Why do natural and laboratory swarms behave so differently? One possible explanation is that external perturbations are responsible for the emergence of critical behavior \cite{vaa20}. However, experiments designed to probe this hypothesis should involve substantially larger laboratory swarms under controlled external perturbations than those studied so far. A successful theory should incorporate interactions between midges and explain how correlations give rise to the critical behavior observed in natural swarms, while also accounting for the characteristic properties of laboratory swarms in the absence of external perturbations. In this paper, we have presented numerical evidence that the anisotropic HCVM reproduces the critical behavior observed in the wild and have explored whether the same model can also account for the properties of laboratory swarms in the absence of external disturbances. The extent to which it succeeds is discussed below.

Laboratory experiments provide probability density functions (PDFs) for the vertical and horizontal positions, velocities, and accelerations of individual midges. Using the anisotropic HCVM, we have calculated these PDFs for the first time in the critical region, obtaining the results shown in Figures \ref{fig6} - \ref{fig8}. The PDFs of the vertical and horizontal velocities are well approximated by $q$-Gaussians with compact support ($q<1$). They differ qualitatively from the PDFs observed in laboratory swarms in the absence of external perturbations and therefore constitute predictions that can be tested against future observations of natural swarms. Probability density functions characterize the stationary invariant measure sampled by the swarm but, by themselves, do not determine the underlying dynamics. Different microscopic dynamics (including independent-particle, chaotic, or correlated many-body systems) may produce similar stationary PDFs. Discriminating among competing theories therefore requires dynamical observables such as correlation functions, critical exponents, and scaling laws.

We have also explored whether the PDFs observed in laboratory swarms correspond to a different region of the HCVM parameter space. Our search was guided by three experimental observations: (i) laboratory swarms contain relatively few insects ($N<100$); (ii) the correlation length remains small over the observed range of swarm RMS radii; and (iii) the position, velocity, and acceleration PDFs are close to Gaussians with long tails. In our quest, we explored two complementary strategies: 
\begin{itemize}
\item Strategy 1: Consider an anisotropic HCVM with zero interaction radius, implying a vanishing correlation length, and explore regions of the phase diagram where the PDFs have kurtoses close to 3, the Gaussian value. We obtained reasonable velocity and acceleration PDFs, but the position PDF was stratified in the horizontal rather than the vertical direction, contrary to observations of small laboratory swarms. Adjusting the ratio between the vertical and horizontal confinement produced the correct stratification but, again contrary to experiment, the effective spring constants increased with insect number.
\item Strategy  2: Remain within the extended critical region associated with the scale-free chaos transition, but move away from the critical line separating chaotic and nonchaotic attractors so that the PDFs approach Gaussian forms. For small $N$, this region is relatively broad. We obtained effective spring constants that decrease with RMS radius, an appropriate susceptibility power law, and PDFs with long tails. However, the resulting polarization was somewhat larger than observed. 
\end{itemize}
These results suggest that the anisotropic HCVM with a fixed ratio between the vertical and horizontal confinement is too restrictive to reproduce the full range of laboratory observations. A natural extension is to explore the three-dimensional phase diagram spanned by the noise amplitude and the horizontal and vertical confinement strengths. 

Beyond the immediate applications discussed here, understanding the scale-free chaos transition of the HCVM in the thermodynamic limit remains an outstanding theoretical challenge. Progress in this direction may eventually provide a deeper understanding of the critical behavior reviewed in this article.
 
\bigskip




{\bf Acknowledgments.}\\ This work has been supported by the FEDER/Ministerio de Ciencia, Innovaci\'on y Universidades -- Agencia Estatal de Investigaci\'on (MCIN/ AEI/10.13039/501100011033) grant PID2024-155528OB-C22.

\end{document}